\documentclass[amsmath,superscriptaddress, amssymb,aps,prd,reprint,longbibliography,nofootinbib,notitlepage,preprintnumbers,10pt]{revtex4-2}
\usepackage{natbib}

\RequirePackage[colorlinks,citecolor=blue,urlcolor=blue,linkcolor=blue]{hyperref}
\usepackage{graphicx}
\usepackage{enumerate}
\usepackage[shortlabels]{enumitem}
\usepackage{url}
\usepackage{xcolor}
\usepackage{graphicx}
\usepackage{setspace}
\usepackage{dutchcal}
\usepackage{multirow}
\usepackage{tabularx}
\usepackage{scalerel}
\usepackage{braket}
\usepackage{tikz}
\usepackage{tikz-cd}
\usetikzlibrary{svg.path}
\usetikzlibrary{decorations.markings}
\usepackage{bbm}
\newcommand{\rd}{{\rm d}}

\newcommand{\e}{{\rm e}}

\usepackage{moreenum}
\usepackage[utf8]{inputenc}

\labelformat{section}{Section #1} 
\labelformat{subsection}{Section #1} 
\labelformat{subsubsection}{Section #1}
\labelformat{subsubsubsection}{Section #1}
\labelformat{equation}{Eq.~(#1)} 
\labelformat{figure}{Fig.~#1} 
\labelformat{subfigure}{Fig.~\thefigure#1} 
\labelformat{table}{Tab.~#1} 
\labelformat{appendix}{Appendix #1}

\definecolor{orcidlogocol}{HTML}{A6CE39}
\tikzset{
  orcidlogo/.pic={
    \fill[orcidlogocol] svg{M256,128c0,70.7-57.3,128-128,128C57.3,256,0,198.7,0,128C0,57.3,57.3,0,128,0C198.7,0,256,57.3,256,128z};
    \fill[white] svg{M86.3,186.2H70.9V79.1h15.4v48.4V186.2z}
                 svg{M108.9,79.1h41.6c39.6,0,57,28.3,57,53.6c0,27.5-21.5,53.6-56.8,53.6h-41.8V79.1z M124.3,172.4h24.5c34.9,0,42.9-26.5,42.9-39.7c0-21.5-13.7-39.7-43.7-39.7h-23.7V172.4z}
                 svg{M88.7,56.8c0,5.5-4.5,10.1-10.1,10.1c-5.6,0-10.1-4.6-10.1-10.1c0-5.6,4.5-10.1,10.1-10.1C84.2,46.7,88.7,51.3,88.7,56.8z};
  }
}

\newcommand\orcidicon[1]{\href{https://orcid.org/#1}{\mbox{\scalerel*{
\begin{tikzpicture}[yscale=-1,transform shape]
\pic{orcidlogo};
\end{tikzpicture}
}{|}}}}

\usepackage{hyperref}
\usepackage{etoolbox}
\makeatletter
\patchcmd{\frontmatter@RRAP@format}{(}{}{}{}
\patchcmd{\frontmatter@RRAP@format}{)}{}{}{}
\renewcommand\Dated@name{}
\makeatother
\renewcommand{\selectlanguage}[1]{}
\usepackage{bm}
\usepackage{xcolor}
\usepackage{bm}
\usepackage{slashed}
\usepackage{xparse}
\usepackage{dsfont}
\usepackage[mathscr]{euscript}
\usepackage{float}
\usepackage{etoolbox}
\patchcmd{\section}
  {\centering}
  {\raggedright}
  {}
  {}
\patchcmd{\subsection}
  {\centering}
  {\raggedright}
  {}
  {}
\DeclareUnicodeCharacter{2212}{-}
\allowdisplaybreaks

\begin{document}
\preprint{YITP-26-159, RESCEU-30/26, IPMU26-0038}
\title{Cosmological correlators and kinematic flow on complex Hartle-Hawking saddles}

\author{Vikramaditya Mondal \orcidicon{0000-0002-8565-5334}}
\email{vikramaditya.academics@gmail.com}
\affiliation{Yukawa Institute for Theoretical Physics, Kyoto University, 606-8502, Kyoto, Japan}

\author{Shinji Mukohyama \orcidicon{0000-0002-9934-2785}}
\email{shinji.mukohyama@resceu.s.u-tokyo.ac.jp}
\affiliation{Research Center for the Early Universe (RESCEU), Graduate School of Science, The University of Tokyo, Hongo 7-3-1, Bunkyo-ku, Tokyo 113-0033, Japan}
\affiliation{Center for Gravitational Physics and Quantum Information, Yukawa Institute for Theoretical Physics, Kyoto University, 606-8502, Kyoto, Japan}

\affiliation{Kavli Institute for the Physics and Mathematics of the Universe (WPI), The University of Tokyo Institutes for Advanced Study (UTIAS), The University of Tokyo, Kashiwa, Chiba 277-8583, Japan}

\begin{abstract}
    The Lorentzian gravitational path integral over homogeneous and isotropic metrics in a closed universe with positive cosmological constant is dominated by complex saddle points, such as the Hartle-Hawking saddles. We ask how to compute cosmological correlators on such complex backgrounds, and how they respond to the shifts of the kinematic variables---the kinematic flow---without tracking the bulk time evolution explicitly. We, specifically, study a conformally coupled scalar test field with a cubic interaction around these saddles. We find that the wave function coefficients can be written as discrete spectral sums with the summands containing the corresponding coefficients in the Einstein static universe, which are obtainable in analogy with the polytopes program for flat spacetime. Whereas the wave function coefficients of the flat FLRW case are given by integrals over shifted kinematics, the discrete spectrum on the three-sphere for Hartle-Hawking saddles produces sums. We identify the coefficients, which depend on the final hypersurface through a phase fixed by its size, with Srivastava-Daoust multi-variable hypergeometric functions. They satisfy recursion relations for any number of sites, under which the sums collapse into rational functions. For the one- and two-site Feynman-Witten tree diagrams, we derive difference equations in the kinematic variables which close on a finite set of sums, giving a discrete kinematic flow. The transformation matrices are upper triangular with rational elements, and satisfy a consistency condition which makes the flow path independent. We then obtain the two-, three- and four-point cosmological correlators. At the late-time conformal boundary, we derive an explicit difference equation for the two-site wave function coefficient.
\end{abstract}

\maketitle
\section{Introduction}
Our universe has structures on large scales which originated due to microscopic quantum fluctuations during inflation. The quantum mechanical origin of these fluctuations is supposed to leave imprints on the late-time structure in terms of correlation functions, making their study one of the central goals in present-day theoretical cosmology \cite{maldacena_non-gaussian_2003,anninos_late-time_2015,arkani-hamed_cosmological_2017,baumann_cosmological_2021,arkani-hamed_differential_2024,goodhew_cosmological_2021,cespedes_time_2021,goodhew_cutting_2021,jazayeri_locality_2021,hillman_differential_2022,meltzer_inflationary_2021,bonifacio_graviton_2023,salcedo_analytic_2022,stefanyszyn_cosmological_2024,maldacena_graviton_2011,mata_cmb_2013,creminelli_conformal_2012,assassi_soft_2012,kundu_constraints_2015,arkani-hamed_cosmological_2015,arkani-hamed_cosmological_2020,baumann_cosmological_2020,sleight_mellin_2020,sleight_bootstrapping_2020,pajer_building_2021,pietro_analyticity_2022,hogervorst_towards_2023,mcfadden_holography_2010,bzowski_holography_2013,baumann_snowmass_2022,benincasa_amplitudes_2022,benincasa_observables_2024,lee_records_2024,albayrak_review_2026,arkani-hamed_kinematic_2025,Baumann:2025qjx,baumann_linking_2022,ghosh_conformal_2014,chakraborty_cosmological_2025,das_reconstructing_2026,das_single-cut_2025,belrhali_laplace_2026}. In this paper, we are particularly interested in how the initial quantum gravitational state of the universe---the Hartle-Hawking state---affects the cosmological correlators. More specifically, we study a toy model: a conformally coupled scalar field on the complex background geometries which dominate the Lorentzian gravitational path integral over homogeneous and isotropic metrics of closed universe with positive cosmological constant.

In a model of cosmology, although everything is described as a time evolution problem starting from some reasonable initial conditions, we can only observe the result of the evolution at some final hypersurface, e.g., the last-scattering surface of the Cosmic Microwave Background Radiation. Thus, it is important to express the cosmological correlators as a function of their kinematic variables. It has been shown that these functions can be determined by a set of self-consistent differential equations called the kinematic flow \cite{arkani-hamed_differential_2024,arkani-hamed_kinematic_2025}. In this description, the bulk time evolution does not have to be explicitly described as long as the set of differential equations for the kinematic flow can be deduced from a simple set of rules. At its core, this program is in line with the cosmological bootstrap program where the correlation functions are described as a function of the boundary data only.

In \cite{arkani-hamed_differential_2024}, it was shown that the wave function coefficients of a conformally coupled field in FLRW spacetime can be represented as integrals where the integrands are proportional to corresponding coefficients in flat spacetime. This is possible due to the fact that under field redefinitions, the action of a conformally coupled field in FLRW spacetime assumes the form of a scalar field action in flat spacetime with a time-dependent coupling ``constant''. Further, the time-dependent coupling factor can be represented as an integral in frequency space. As a result, the FLRW wave function coefficients can be represented as integrals over the same in flat spacetime.

In the present case, as the spatial geometry is closed, this correspondence exists between Hartle-Hawking (HH) saddle point geometries and the Einstein Static Universe (ESU).

In quantum cosmology, the quantum state of the universe is described by the ``wave function of the universe'', $\Psi[h]$, which satisfies the Wheeler-DeWitt equation $\hat{\mathscr{H}}[h] \Psi[h]=0$ \cite{DeWitt:1967yk}, where $\hat{\mathscr{H}}[h]$ is the operator corresponding to the Hamiltonian constraint in the Arnowitt, Deser, and Misner (ADM) formulation of General Relativity \cite{Arnowitt:1962hi}. Here, $h$ represents both the induced metric $h_{ij}$ on spatial hypersurfaces of constant time and any other matter degrees of freedom, for example, a scalar field. $\Psi[h]$ is a function(al) living in the configuration space of all 3-geometries (modulo spatial diffeomorphisms) and field configurations called the \textit{superspace} \footnote{This naming has no connection to theories with supersymmetry.}. Even though the dimension of the superspace is infinite, for cosmological scenarios, the symmetry of the early universe, i.e., spatial homogeneity, reduces the number of degrees of freedom and results in a finite-dimensional configuration space, called the \textit{minisuperspace}. In the minisuperspace setup, the Wheeler-DeWitt equation can be solved in many interesting cases with prescriptions for boundary conditions \cite{vilenkin_boundary_1986}.

It is also possible to construct a path integral formalism as an equivalent alternative to the canonical formalism \cite{Teitelboim:1981ua,Teitelboim:1983fh,PhysRevD.28.310,PhysRevD.28.297,Barvinsky:1986qn,Barvinsky:1986bt,Kuchar_Skeletonizations,hartle1984path,PhysRevD.34.2323,Halliwell:1988wc,Halliwell:1990qr,Hartle:1983ai,Feldbrugge:2017kzv,DiTucci:2019dji,di2019noprescription}, such that the wave function is defined as a path integral: $\Psi[g]=\int \mathscr{D}{}g^{(4)} \, \e^{\frac{i}{\hbar}S[g^{(4)}]}$, where the integral should be only over metrics that are distinct modulo diffeomorphisms. Extra care needs to be taken to incorporate the Hamiltonian and momentum constraints utilizing the BRST formalism \cite{Halliwell:1988wc,Halliwell:1990qr}. In earlier works in quantum cosmology, the wave function of the universe was often defined with the Euclidean path integral, rather than its Lorentzian definition (see, for example, \cite{hartle_wave_1983,PhysRevD.39.2206}). Apart from the possible issue of the non-equivalence of summing over Euclidean versus Lorentzian geometries, the Euclidean path integral is also not well-defined mathematically, as these integrals may fail to converge \cite{gibbons_einstein_1977,gibbons_path_1978}. As a consequence of this, a Lorentzian path integral needs to be computed, and in the semi-classical approximation, the path integral can be evaluated by means of the steepest-descent approximation utilizing Picard-Lefschetz theory \cite{feldbrugge2017lorentzian}. The complex saddle point geometries that dominate the steepest descent approximation of the Lorentzian path integral for a homogeneous and isotropic universe with a positive cosmological constant serve as the background spacetime for this current paper.

Evaluation of the path integral requires information regarding the initial hypersurface, which is not known. In this paper, we particularly consider the `no-boundary' proposal due to Hartle and Hawking \cite{hartle_wave_1983}. This proposal originally purported to have resolved the initial singularity of the universe by going to the Euclidean time domain. In that case, the hyperbolic Lorentzian de Sitter spacetime becomes a 4-sphere, and the size of the universe can be smoothly taken to vanish at the initial (Euclidean) time. The wave function is then described as a Euclidean path integral in which one sums over all possible regular geometries that start with zero size but end in the same final boundary hypersurface, which can be identified with the beginning of a Lorentzian signature de Sitter universe. Therefore, the geometries summed over in the path integral have only one boundary, the final hypersurface, but no initial boundary---no-boundary geometries. However, as there are sufficient objections to the Euclidean gravitational path integral, new Lorentzian techniques were applied to re-derive Hartle-Hawking's results \cite{feldbrugge2017lorentzian,feldbrugge2017no}. Two major issues were noticed in these works: (i) the wave function comes with a prefactor usually ascribed to Vilenkin's definition of the path integral \cite{vilenkin_quantum_1984,vilenkin_quantum_1985,PhysRevD.37.888,vilenkin2018tunneling,vilenkin_boundary_1986} and not what Hartle-Hawking originally found \cite{Feldbrugge:2017fcc,Feldbrugge:2017mbc,honda_dimensional_2026}, and (ii) the saddle points picked up by the Lorentzian path integral were unstable against gravitational perturbations \cite{feldbrugge2017no}. These observations created a controversy (see, for example, \cite{feldbrugge2017lorentzian,feldbrugge2017no,dorronsoro2017real,dorronsoro2018damped,vilenkin2018tunneling,feldbrugge2018no,feldbrugge2018inconsistencies,vilenkin2019tunneling}). Later, it was proposed that to avoid running into the problem of uncontrolled perturbations around the saddle point geometries, the original idea of the creation of the universe from `nothing' or an initial zero-size universe might have to be traded off in favor of other suitable initial boundary conditions (such as Neumann and Robin conditions) that lead to stable saddle points \cite{di2019no,di2019noprescription,rajeev_wave_2022}, or more recently, prescriptions defining the integrals with regulators or a small initial boundary \cite{yamada_new_2026,yamada_small-throat_2026}, among other ideas on initial conditions \cite{ailiga_lorentzian_2024,ailiga_boundary_2025,mallik_ir_2026,ailiga_initial_2026}. In the new Lorentzian approach, the wave function of the universe is defined with initial Neumann or Robin boundary conditions, which requires one to add appropriate boundary terms to the action such that the variational principle with respect to these boundary conditions remains consistent \cite{di2019noprescription}.

Fluctuations and correlators in a closed universe have also been studied in the context of quantum cosmology. In \cite{Halliwell:1984eu}, the perturbations around a closed universe with the no-boundary initial condition were analysed by expanding them in the harmonics on $\mathbb{S}^3$. More recently, the fluctuations of the Hartle-Hawking and tunneling wave functions were studied in the Hamiltonian formalism \cite{deAlwis:2018sec}, and the density perturbations and primordial non-Gaussianities in a closed universe were computed in \cite{Cespedes:2021oso}. Further, the scattering amplitudes of scalars and gluons in global de Sitter spacetime, whose spatial slices are three-spheres, have been studied in \cite{Taylor:2024vdc,Taylor:2025spp,Lindsay:2025chz,Taylor:2026ase}. See also \cite{Chakraborty:2023yed,chakraborty_cosmological_2025} for the Hilbert space and the cosmological correlators in de Sitter quantum gravity.

The goal of this paper is to rigorously evaluate cosmological correlators on these complex Hartle-Hawking saddle point geometries and map their kinematic dependence. However, because we are dealing with a closed spatial topology ($\mathbb{S}^3$) rather than flat space ($\mathbb{R}^3$), the momentum modes of the test field are inherently discrete. Consequently, the continuous energy integrals in the flat-space cosmological polytope program are replaced by discrete spectral sums over the Einstein Static Universe mode functions. This discreteness implies that the continuous differential equations governing standard kinematic flow are replaced, and the kinematic dependence of the correlators is governed by discrete shift transformations.  We demonstrate, specifically for one- and two-site Feynman-Witten tree diagrams that this transformation forms a closed, upper triangular matrix algebra which satisfies discrete integrability condition. Thus, the continuous kinematic flow of flat FLRW spacetime gets translated into an exactly solvable, discrete matrix flow on the complex Hartle-Hawking saddle.

Our main results can be summarized as follows. (i) The wave function coefficients on the Hartle-Hawking saddles are given by phase-weighted discrete sums over the shifted coefficients on the Einstein static universe (see \ref{eq:HH-sum-general}), which we identify with the Srivastava-Daoust multi-variable hypergeometric series at all orders. (ii) These coefficients satisfy recursion relations at all orders with any number of sites (see \ref{eq:recursion-V}), under which all the sums collapse into rational functions of the kinematic variables. (iii) For the one- and two-site diagrams, the shifts of the kinematic variables close on finite sets of two and six sums, respectively, giving the discrete kinematic flow in \ref{eq:flow-one-site} and \ref{eq:flow-two-site}. The transformation matrices are upper triangular, with elements that are rational functions of the kinematic variables, and they satisfy the consistency condition \ref{eq:consistency}, which makes the flow independent of the path in the discrete kinematic space. (iv) We obtain the two-, three- and four-point correlation functions in \ref{eq:two-point}, \ref{eq:three-point} and \ref{eq:four-point}, where the generalized Gaunt coefficients on $\mathbb{S}^3$ replace the momentum conserving Dirac delta functions of flat space. Since the correlators depend only on the real parts of the wave function coefficients, they are the same on both the Hartle-Hawking saddles. (v) At the late-time conformal boundary, all the shifted coefficients on the Einstein static universe enter the sums with the same weight, so that the sums become discrete counterparts of the integral representation in de Sitter spacetime. We give the transformation matrices in this limit in \ref{eq:one-site-late} and \ref{eq:two-site-late}, and show that the two-site coefficient entering the four-point function is real and satisfies the explicit difference equation \ref{eq:late-time-psi2}.

The paper is organized as follows. In \ref{sec:saddles}, we briefly discuss the complex Hartle-Hawking saddles which dominate the Lorentzian path integral with the initial Neumann and final Dirichlet boundary conditions, and the complex contour of the conformal time on these saddles. In \ref{sec:ESU-vs-HH}, we show that a conformally coupled scalar field on these saddles is equivalent to the same field on the Einstein static universe with a time-dependent coupling, and express this time dependence as a discrete sum over frequencies. In \ref{sec:path-integral}, we set up the path integral over the field and define the wave function coefficients. In \ref{sec:sums}, we compute the wave function coefficients on the Einstein static universe, express the coefficients on the Hartle-Hawking saddles as discrete sums over them, and identify these sums with multi-variable hypergeometric series. In \ref{sec:kinematic-flow}, we derive the recursion relations at all orders, and the difference equations and the discrete kinematic flow for the one- and two-site diagrams, which is the main result of this paper. In \ref{sec:convergence-late-time}, we discuss the convergence of the sums and their late-time limit, where we give the transformation matrices and an explicit difference equation for the two-site coefficient. Then, in \ref{sec:correlators}, we obtain the cosmological correlators, the power spectrum and the bispectrum. We conclude in \ref{sec:conclusion}. The background equations for the saddles, the construction of the bulk-to-bulk propagator and the evaluation of the time integrals on the Einstein static universe, and the derivation of the transformation matrices for the two-site diagram and the proof of the late-time difference equation are given in the appendices.

\section{Complex Hartle-Hawking saddles}
\label{sec:saddles}
For the Lorentzian path integral approach \cite{feldbrugge_lorentzian_2017}, it is convenient to start with the following form of the metric
\begin{align}\label{eq:metric}
    \rd s^2 = - \frac{N^2(t)}{q(t)} \rd t^2 + q(t) \rd\Omega_{3}^2,
\end{align}
where $q(t)\equiv a^2(t)$ is the (squared) scale factor, $\rd\Omega_{3}^2$ is the metric on a unit 3-sphere, and $N(t)$ is the lapse function. The above choice of the metric is motivated from the fact that in these coordinates the Einstein-Hilbert action takes a convenient form, with the minisuperspace potential being only a linear function of $q$, allowing us to perform the path integral over $q$ exactly before integrating over the lapse function $N$.

In the quantum theory, the notion of equations of motion and classical trajectories as solutions to them no longer applies. Instead, we have to assign a wave function to the universe defined as follows
\begin{equation}
\Psi_{\sf initial}[q_{\sf f},{\phi}_{\sf f}] = \int_\mathcal{C}\rd N \int_{\sf initial}^{(q_{\sf f},{\phi}_{\sf f})} \mathcal{D}[q] \mathcal{D}[\phi] ~ e^{i S[N,q,\phi]/\hbar} .
\end{equation}
Here, the action, $S[N,q,\phi]=S_{\text{E-H}}[q,N]+\mathcal{S}[\phi,q,N]$, contains both the gravitational action and the action of a field $\phi$. The wave function is given as a function(al) of $(q_{\sf f},\phi_{\sf f})$ which are the (squared) scale factor and field configuration on the final 3-hypersurface. The integration over the lapse function on the full line $(-\infty,\infty)$ is carried out along a complex contour $\mathcal{C}$ ensuring the convergence of the oscillatory integrand and depends on the choice of initial and final boundary conditions. The choice of this contour is determined by the application of Picard-Lefschetz theory \cite{feldbrugge_lorentzian_2017}. The initial condition for the quantum gravitational path integral does not come from any known physics, and it is argued that this should be a new principle of physics in its own right. As such we have to rely on proposals for the initial conditions, and these have been a subject of fierce debate in the past few decades as we discussed above. In this paper we shall discuss the case of the Hartle-Hawking no-boundary proposal.

To obtain the no-boundary wave function, one has to sum over all geometries that start with vanishing size $q(0)=0$. However, it has been shown that the saddle point geometry dominating the Lorentzian path integral with this initial Dirichlet condition leads to unstable \textit{inverse Gaussian} perturbations \cite{Feldbrugge:2017fcc,Feldbrugge:2017mbc}. To cure this issue and recover a stable no-boundary wave function, alternative initial conditions have been suggested \cite{di2019no,di2019noprescription,rajeev_wave_2022,yamada_new_2026,yamada_small-throat_2026}. Here, we shall work with the simplest of such prescriptions, where an initial Neumann and final Dirichlet boundary condition is chosen \cite{di2019noprescription}.

Ignoring the backreaction of the matter field on the geometry we can approximate the wave function as
\begin{align}\label{eq:path-integral-saddle-approx}
    \Psi_{\sf initial}[q_{\sf f},\phi_{\sf f}] \simeq \sum_{\sigma} \e^{i S_{\rm cl}(\bar{q}_{\sigma},N_{\sigma})/\hbar}\int\mathcal{D}[\phi] \e^{i \mathcal{S}[\phi,\bar{q}_{\sigma},N_{\sigma}]/\hbar},
\end{align}
where $\sum_{\sigma} \e^{i S_{\rm cl}(\bar{q}_{\sigma},N_{\sigma})/\hbar}$ is the saddle point approximation to the background Lorentzian gravitational path integral. $S_{\rm cl}$ is the Einstein-Hilbert action (without matter field) evaluated for $\bar{q}$, which is the solution to the background equation of motion $\delta S_{\text{E-H}}/\delta q|_{\bar{q}} =0$, with the initial Neumann and final Dirichlet boundary conditions; and also on $N_{\sigma}$ which are the saddle points of the function $S_{\rm cl}(\bar{q},N)$ defined as $\partial S_{\rm cl}/\partial N|_{N_{\sigma}} = 0$. And $\mathcal{S}[\phi,\bar{q}_{\sigma},N_{\sigma}]$ is the action of a test scalar field on the saddle point background geometry $(\bar{q}_{\sigma},N_{\sigma})$. In this paper, we choose the test field to be a conformally coupled scalar field with a cubic interaction, and complex saddle point geometries $(\bar{q}_{\sigma},N_{\sigma})$ correspond to the Hartle-Hawking no-boundary proposal \cite{di2019noprescription}.

\begin{figure}
    \centering
    \includegraphics[width=0.95\linewidth]{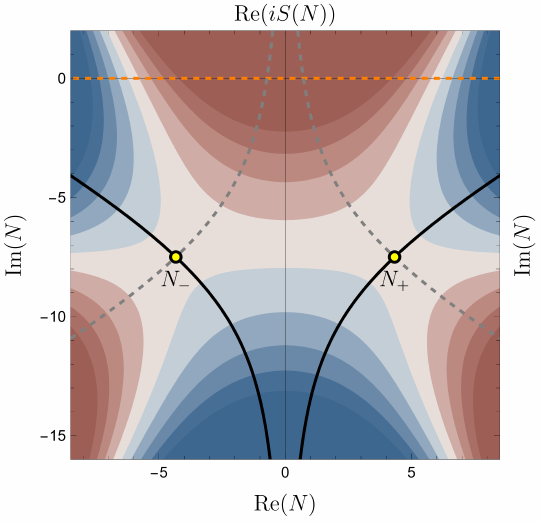}
    \caption{Contour plot of ${\rm Re}[iS(\bar{q},N)]$ or the Morse function; in the bluer regions the Morse function has large negative values (convergent region), and in the redder regions the Morse function has large positive values (divergent region). $N_{\pm}$ are the saddle points defined in \ref{eq:saddles_inf_neu_diri}. The orange dashed line, along ${\rm Im}(N)=0$, is the original contour of integration, which is deformed along the steepest descent contours or the Lefschetz thimbles (solid black lines) that pick up the two saddle points contributing to the lapse integral. The steepest ascent lines from the saddle points are displayed by gray dashed lines. For this plot we have chosen $q_{1}=10$, $\Lambda=2/5$.}
    \label{fig:saddles}
\end{figure}

The action for the background geometry with the appropriate boundary terms, the equations of motion, and their solution satisfying the initial Neumann condition $\dot{\bar{q}}(0)/2N=\pi_{0}$ and the final Dirichlet condition $\bar{q}(1)=q_{1}$ are summarized in \ref{app:saddles}.
The total action, $S=S_{\text{E-H}}+S_{\partial\mathscr{M}}+S_{B}$, evaluated at this classical solution $S_{\rm cl}(\bar{q},N)$ has saddle points in the complex lapse plane at
\begin{align}
\label{eq:saddles_inf_neu_diri}
N_{\pm} = \frac{3}{\Lambda} \left(-\pi_{0} \pm \sqrt{\frac{\Lambda}{3}q_{1} - 1} \right).
\end{align}
Further, one chooses the initial momentum to be
\begin{align}\label{eq:no-boundary-cond}
    \pi_{0} = + i, 
\end{align}
which makes saddle point geometry
\begin{align}\label{eq:saddle_geometry}
    \bar{q}_{\sigma}(t) & = \frac{\Lambda}{3} N^2_{\sigma} t^2 + 2 i N_{\sigma} t + q_{1} - \frac{\Lambda}{3} N^2_{\sigma} - 2 i N_{\sigma}, ~~~~ \sigma=\pm.
\end{align}
to be a no-boundary geometry, that is $\bar{q}_{\sigma}(0)=0$. The saddle point approximation to the lapse integral picks up contributions from the saddle points $N_{\pm}$ after deforming the original integration contour along the real contours into the steepest descent contours from the saddle points or the Lefschetz thimbles (see \ref{fig:saddles}). The sum in the \ref{eq:path-integral-saddle-approx} is over these two saddle points.

Our goal in this paper is to compute the wave function of a conformally coupled scalar (test) field
\begin{align}\label{eq:scalar-PI}
    \Psi_{\rm HH}[\Phi(\boldsymbol{x}),q_{1}] = \int_{\phi(0)=0}^{\phi(1)=\Phi(\boldsymbol{x})}\mathcal{D}[\phi] \e^{i \mathcal{S}[\phi,\bar{q}_{\sigma},N_{\sigma}]/\hbar},
\end{align}
on the background Hartle-Hawking no-boundary saddle geometries $(\bar{q}_{\sigma},N_{\sigma})$ defined by \ref{eq:saddles_inf_neu_diri}, \ref{eq:no-boundary-cond}, \ref{eq:saddle_geometry}. It is clear that even though the saddle geometries $(\bar{q}_{\sigma},N_{\sigma})$ are complex, the (squared) scale factor on the final hypersurface $q_{1}$ is real. 

Since we will be dealing with a conformally coupled field, it is convenient to consider the FLRW metric \ref{eq:metric} in the conformal form
\begin{align}
    \rd s^2 = - \frac{N^2(t)}{q(t)} \rd t^2 + q(t) \rd\Omega_{3}^2 = q(\eta)\left[-\rd \eta^2 + \rd\Omega_{3}^{2} \right],
\end{align}
where we have defined the conformal time as
\begin{align}
    \rd \eta = \frac{N(t)}{q(t)}\rd t .
\end{align}
For Lorentzian no-boundary geometry defined by \ref{eq:saddles_inf_neu_diri}, \ref{eq:no-boundary-cond}, \ref{eq:saddle_geometry}, the conformal time $\eta$ is related to the coordinate time $t$ as
\begin{align}
    \eta_{\pm} = N_{\pm} \int \frac{\rd t}{\frac{\Lambda}{3} N_{\pm}^2 t^2 + 2 N_{\pm} \pi_{0} t + q_{1} - \frac{\Lambda}{3} N_{\pm}^2 - 2 N_{\pm} \pi_{0}},
\end{align}
which simplifies to
\begin{align}
    \eta_{\pm} = \arctan\left(\frac{\Lambda}{3}N_{\pm} t+ i\right) + C .
\end{align}
It is convenient to work with the choice $C = 0$. Let us consider the boundaries of the conformal time corresponding to the interval $t\in[0,1]$:
\begin{subequations}\label{eq:conformal-time-range}
\begin{align}
    t = 1, ~~~~ \eta_{\pm} & \to \eta_{\rm f} \equiv \arctan\left(\pm\sqrt{\frac{\Lambda}{3}q_{1}-1}\right) , \\
    t \to 0, ~~~~ \eta_{\pm} & \to +i\infty + \frac{\eta_{\rm f}}{2} .
\end{align}    
\end{subequations}
Recall that in flat FLRW case, the choice of the Bunch-Davies vacuum requires an $i\epsilon$ prescription $\eta\to \eta(1-i\epsilon)$ so that in the limit $\eta\to-\infty$, the correct mode functions $\e^{ik\eta}$ are chosen. Here, in the case of Hartle-Hawking saddle geometries, the conformal time runs along a complex contour and automatically selects the Bunch-Davies vacuum without needing any $i\epsilon$ prescription.

Since the conformal time is complex, the reality of the endpoint, or the conformal time at the hypersurface $t=1$, depends on the corresponding scale factor $q_{1}$. If $q_{1}>3/\Lambda$, which corresponds to a universe in the classically allowed region, then the conformal time at $t=1$ endpoint is real, and is either positive or negative for the $N_{+}$ or $N_{-}$ saddle points, respectively. If we consider a universe nucleating with the final size close to the de Sitter throat $q_{1} \to 3/\Lambda$, from above corresponds to the conformal time $\eta\to 0^{\pm}$, for $N_{+}$ and $N_{-}$, respectively.

When the final time hypersurface $t=1$ corresponds to a very large universe $q_{1}\to\infty$, the conformal time approaches $\eta_{\pm}\to\pm\frac{\pi}{2}$. Thus, the saddle point geometries $(\bar{q}_{\pm},N_{\pm})$ both originate with no-boundary $\bar{q}_{\pm}(+i\infty+\eta_{\rm f}/2)=0$, and the late time hypersurface of $\bar{q}_{+}$ corresponds to the late-time conformal boundary of de Sitter spacetime in global coordinates at $\eta_{+}=+\frac{\pi}{2}$, and similarly late time hypersurface of $\bar{q}_{-}$ corresponds to conformal boundary at $\eta_{-}=-\frac{\pi}{2}$. 
\section{Conformal fields on complex Hartle-Hawking saddles}
\label{sec:ESU-vs-HH}
We start by considering a conformally coupled scalar field with the following action
\begin{align}
\mathcal{S} = & \int {\rm d}^4 x \sqrt{-g}\left[-\frac{1}{2} g^{\mu \nu} (\partial_{\mu} \phi) (\partial_{\nu} \phi) - \frac{1}{12} R \phi^2 + \frac{\lambda_{p}}{p!} \phi^{p} \right] .
\end{align}

In the conformal time coordinates, the action for the scalar field takes the form
\begin{align} \label{eq:conformal_action}
\mathcal{S} = & \int {\rm d}\eta \rd \boldsymbol{x} \sqrt{h} \left[\frac{1}{2} q (\phi')^2 - \frac{1}{2}  q (\boldsymbol{\nabla}_{\mathbb{S}^3}\phi)^2 - \frac{1}{12} q^2 R \phi^2 \right. \nonumber\\
& \left.+ \frac{\lambda_{p}}{p!}q^2 \phi^{p} \right].
\end{align}
Here, prime denotes derivative with respect to the conformal time $\eta$, and the Ricci scalar of the spacetime has the form
\begin{align}
    R = \frac{3 q''}{q^2}-\frac{3 q'^2}{2 q^3}+\frac{6}{q} .
\end{align}
Further, $h$ and $\boldsymbol{\nabla}_{\mathbb{S}^3}$ are the determinant of the metric and covariant derivative on $\mathbb{S}^3$. Under the field redefinition
\begin{align}\label{eq:field-rescaling-1}
    \phi \to \frac{\phi}{\sqrt{q(\eta)}},
\end{align}
the above action \ref{eq:conformal_action}, up to a boundary term takes the form
\begin{align}
\mathcal{S} = & \int {\rm d}\eta \rd^3 \boldsymbol{x} \sqrt{h} \left[\frac{1}{2}(\phi')^2 - \frac{1}{2} (\boldsymbol{\nabla}_{\mathbb{S}^3}\phi)^2 - \frac{1}{2}\phi^2 \right. \nonumber\\
& \left. + \frac{\lambda_{p}}{p!}q^{2-\frac{p}{2}} \phi^{p} \right].
\end{align}

On the other hand, the metric on an Einstein static universe has the form
\begin{align}
    \rd s^2 = R_{\mathbb{S}^3}^2\left[-\rd \eta^2 + \rd\Omega_{3}^{2} \right].
\end{align}
Here, $R_{\mathbb{S}^3}$ is the radius of the universe, which has a constant curvature determined by the radius as
\begin{align}
    R = \frac{6}{R_{\mathbb{S}^3}^2} .
\end{align}
The action for a conformally coupled scalar field in this spacetime has the form
\begin{align}
\mathcal{S}_{\rm ESU} = & \int {\rm d}\eta \rd^3 \boldsymbol{x} \sqrt{h} \left[\frac{1}{2}(\phi')^2 - \frac{1}{2} (\boldsymbol{\nabla}_{\mathbb{S}^3}\phi)^2 - \frac{1}{2}\phi^2 \right. \nonumber\\
& \left.+ \frac{\lambda_{p}}{p!} R_{\mathbb{S}^3}^{4-p} \phi^{p} \right] ,
\end{align}
where we have redefined the field to
\begin{align}\label{eq:field-rescaling-2}
    \phi \to \frac{\phi}{R_{\mathbb{S}^3}} .
\end{align}
Therefore, the conformally coupled scalar field on the Hartle-Hawking saddles can be thought of as a conformally coupled field in the Einstein static universe with time-dependent couplings for interactions with the identification
\begin{align}
    \lambda_{p}^{\rm ESU} R_{\mathbb{S}^3}^{4-p} \leftrightarrow q(\eta)^{2-\frac{p}{2}} \lambda_{p}^{\rm HH} .
\end{align}
It is then important to understand the frequency domain representation of the time dependence $q^{2-p/2}(\eta)$ of the coupling factor if we want to express the Hartle-Hawking wave function coefficients as sum over those of the Einstein static universe.

Now, since we have
\begin{align}
\bar{q}_{\pm}(t) = N_{\pm}t\left(\frac{\Lambda}{3} N_{\pm} t + 2 i\right) ,
\end{align}
then the scale factor can be expressed in terms of the conformal time as
\begin{align}
\bar{q}_{\pm}(\eta) = \frac{3}{\Lambda}\left(\tan\eta_{\pm} - i\right)\left(\tan\eta_{\pm} + i\right) =\left( \frac{\sqrt{3/\Lambda}}{\cos\eta_{\pm}}\right)^2.
\end{align}
The time dependence of the coupling factor in the Hartle-Hawking saddles can be written as
\begin{align}
    \lambda^{\rm HH}_{p}(\eta) = & \lambda_{p}^{\rm HH} \left(\frac{\Lambda}{3}\right)^{\frac{p}{2}-2} \cos^{p-4} \eta \nonumber\\
    = & \sum_{k=0}^{\infty}\begin{pmatrix}
        p-4\\
        k
    \end{pmatrix} \lambda_{p}^{\rm HH} \left(\frac{\Lambda}{3}\right)^{\frac{p}{2}-2} (2)^{4-p} \e^{i(2k+4-p)\eta} .
\end{align}
Here, for any real number $\alpha$ and non-negative integer $k$, the binomial coefficients are defined as
\begin{align}
    \begin{pmatrix}
        \alpha\\
        k
    \end{pmatrix} = \frac{\alpha(\alpha-1)\cdots(\alpha-k+1)}{k!} ,
\end{align}
which do not vanish for any $k$ when $\alpha$ is a negative integer, e.g., $\binom{-1}{k}=(-1)^{k}$.
Thus we can write the time dependence as sum of the following form
\begin{align}\label{eq:spectral}
    \lambda_{p}^{\rm HH}(\eta) = \sum_{k=0}^{\infty} \tilde{\lambda}_{p}(\omega_k) \e^{i\omega_k \eta},
\end{align}
where we have defined
\begin{subequations}
\begin{align}
    \omega_{k} & = 2k+4-p , \\
    \tilde{\lambda}(\omega_{k}) & = \begin{pmatrix}
        p-4\\
        k
    \end{pmatrix} \lambda_{p}^{\rm HH} \left(\frac{\Lambda}{3}\right)^{\frac{p}{2}-2} (2)^{4-p} .
\end{align}    
\end{subequations}
Note that, as the spatial geometry is $\mathbb{S}^3$ in the case of the Hartle-Hawking saddles, the energy modes are discrete, and as a result the wave function coefficients depend on discrete kinematics. This is the reason why we re-expressed the time dependence of $\lambda(\eta)$ as a discrete sum, so that we can express, as we will show below, the Hartle-Hawking wave function coefficients as a discrete sum over shifted Einstein static universe coefficients.

Note that for $p=4$ the interaction is conformally invariant and the coupling is time independent, while for $p>4$ the binomial coefficients vanish for $k>p-4$, so that the sum is finite. We will particularly discuss in this paper the case of $p=3$, for which the sum is infinite, and as a result interesting mathematical structures emerge. In this case, the above decomposition of the coupling into discrete modes is valid only for ${\rm Im}(\eta)>0$, which is satisfied along the complex contour of the conformal time on the no-boundary geometries, except at the final hypersurface where $\eta_{\rm f}$ is real. We discuss the convergence of the resulting sums on the final hypersurface in \ref{sec:convergence-late-time}.

Since we are interested in the cubic theory, in this case, the relations simplify to
\begin{align}
    \tilde{\lambda}(\omega_{k} = 2k+1) = 2\lambda_{p}^{\rm HH} \sqrt{\frac{3}{\Lambda}}(-1)^k .
\end{align}
In the following, we consider a theory with cubic interactions and discuss how to compute the wave function coefficients corresponding to this case.

\section{Path integral over the field}
\label{sec:path-integral}
In the following, we will primarily follow the formalism discussed in \cite{benincasa_amplitudes_2022}. We will be working in the perturbative regime with an action of the form
\begin{align}
    \mathcal{S} = \mathcal{S}_{\rm free} + \mathcal{S}_{\rm int},
\end{align}
where $\mathcal{S}_{\rm free}$ is the quadratic part of the action and $\mathcal{S}_{\rm int}$ contains the perturbative interaction; in the present case, we consider $p=3$. We shall split the field into its free part and deviation from thereof due to interaction as 
\begin{align}\label{eq:scalar-flucctuations}
    \phi(x) = \phi_{\circ}(x) + \varphi(x) .
\end{align}
Here, $\phi_{\circ}(x)$ is the solution to the free equation of motion and $\varphi(x)$ is the fluctuation. In the path integral, we will consider a field that initially has a vanishing configuration and has the distribution $\Phi(\boldsymbol{x})$ on the final hypersurface. Then the free and fluctuation components of the field should satisfy the following boundary conditions
\begin{subequations}
\begin{align}
    \phi_{\circ}(+i\infty+\eta_{\rm f}/2,\boldsymbol{x}) & = 0, ~~~~ & \phi_{\circ}(\eta_{\rm f},\boldsymbol{x}) &= \Phi(\boldsymbol{x}) , \label{eq:free-boundary-condition} \\
    \varphi(+i\infty+\eta_{\rm f}/2,\boldsymbol{x}) & = 0, ~~~~ &\varphi(\eta_{\rm f},\boldsymbol{x}) &= 0 .
\end{align}    
\end{subequations}
Here, $\eta_{\rm f}$ is the conformal time at the final hypersurface, whereas the initial conformal time is $+i\infty+\eta_{\rm f}/2$ as discussed above (see, \eqref{eq:conformal-time-range}). For a given scale factor at the final hypersurface $\eta_{\rm f}=\arctan\left(\pm\sqrt{\Lambda q_{1}/3 -1}\right)$ depending on which saddle is being considered. However, for most of the discussion below we will keep the generic $\eta_{\rm f}$ as a placeholder.

The splitting of the field leads to the following splitting of its action up to boundary terms as
\begin{align}
    \mathcal{S} = \mathcal{S}_{\rm free}[\phi_{\circ}] + \mathcal{S}_{\rm free}[\varphi] + \mathcal{S}_{\rm int}[\phi_{\circ},\varphi] ,
\end{align}
Where $\mathcal{S}_{\rm free}[\varphi]$ is the quadratic part of the action with $\varphi(x)$ as the field variable. Then the path integral in \ref{eq:scalar-PI} becomes
\begin{align}
    \Psi_{\rm HH}[\Phi(\boldsymbol{x})] = & \mathscr{N} \e^{i\mathcal{S}_{\rm free}[\phi_{\circ}]} \int_{\varphi(+i\infty+\frac{\eta_{\rm f}}{2},\boldsymbol{x})=0}^{\varphi(\eta_{\rm f},\boldsymbol{x})=0} \mathcal{D}[\varphi] \e^{i\mathcal{S}_{\rm free}[\varphi]} \nonumber\\
    & ~~~~~~~~ \times \e^{+ i\mathcal{S}_{\rm int}[\phi_{\circ},\varphi]},
\end{align}
where the background spacetime is understood to be given by $(\bar{q}_{\sigma},N_{\sigma})$, and we have suppressed the label $q_{1}$. Since the interactions are being treated perturbatively, we can expand the exponential for the interaction part of the action and obtain
\begin{align}\label{eq:expansion}
    \Psi_{\rm HH}[\Phi] =  \e^{i \mathcal{S}_{\rm free}[\Phi]} \sum_{j=0}^{\infty} \frac{i^j}{j!}\langle \mathcal{S}^j_{\rm int}  \rangle [\Phi] ,
\end{align}
where we have defined
\begin{align}
    \langle \mathcal{S}^j_{\rm int}  \rangle [\Phi] = \mathscr{N} \int_{\varphi(+i\infty+\frac{\eta_{\rm f}}{2},\boldsymbol{x})=0}^{\varphi(\eta_{\rm f},\boldsymbol{x})=0} \mathcal{D}[\varphi] \e^{i\mathcal{S}_{\rm free}[\varphi]} \mathcal{S}^j_{\rm int}[\phi_{\circ},\varphi] ,
\end{align}
and the normalization is given by
\begin{align}
    \mathscr{N}^{-1} = \int_{\varphi(+i\infty+\frac{\eta_{\rm f}}{2},\boldsymbol{x})=0}^{\varphi(\eta_{\rm f},\boldsymbol{x})=0} \mathcal{D}[\varphi] \e^{iS_{\rm free}[\varphi]} .
\end{align}
The expansion in \ref{eq:expansion} can be represented diagrammatically \cite{benincasa_amplitudes_2022}, and in the following we shall discuss the one- and two-vertex tree diagrams.

\subsection{Free theory}
Let us first consider the free theory before dealing with the interaction. We take the following \textit{ansatz} for the free field
\begin{align}
\label{phix}
\phi_{\circ}(x)& =\sum_{n \ell m} \phi_{n\ell m}(\eta) Q_{n \ell m}(\bm{x}),
\end{align}
where the $Q_{n\ell m}(\boldsymbol{x})$ are the spherical harmonics on $\mathbb{S}^3$, which satisfy the Helmholtz equation
\begin{align}
\boldsymbol{\nabla}_{\mathbb{S}^3}^{2}Q_{n \ell m}(\chi,\theta,\phi) = -(n^2-1) Q_{n \ell m} (\chi,\theta,\phi) \, ,
\end{align}
where the index $n \in \mathbb{N}$ goes over all the natural numbers \cite{wen_properties_1985,lindblom_scalar_2017,rubin_eigenvalues_1984,higuchi_symmetric_1987}. Note that in the literature it is common to represent the eigenvalues as $L(L+2)$ with $L$ being any non-negative integer including $0$. However, one can choose a parametrization $n=L+1$, such that $n$ goes over all natural numbers.

The harmonics are normalized as follows
\begin{align}\label{space_norm}
\int_{\mathbb{S}^3} {\rm d}^3 \bm{x} \sqrt{h} Q^*_{n \ell m}(\bm{x})Q_{n'\ell'm'}(\bm{x}) = \delta_{nn'}\delta_{\ell \ell'}\delta_{mm'} .
\end{align}
This separation of variables in the ansatz \ref{phix} allows the action to be written as
\begin{align}
    \mathcal{S}_{\rm free}[\phi_{\circ}]  = \sum_{n\ell m} \mathcal{S}_{n\ell m} ,
\end{align}
where the summand specified by the quantum numbers $(n,\ell,m)$ has the form 
\begin{align}\label{eq:action_scalar_eigen} 
\mathcal{S}_{n\ell m}& = \frac{1}{2}\int_{\mathcal{C}(+i\infty+\frac{\eta_{\rm f}}{2},\eta_{\rm f})} {\rm d}\eta \left(\phi'_{n\ell m} {\phi'}_{n\ell m}^* - n^2 \phi_{n\ell m}\phi_{n\ell m}^*\right) \nonumber\\
& = \frac{1}{2} \phi'_{n\ell m}\phi_{n\ell m}^*\Bigg|_{+i\infty+\frac{\eta_{\rm f}}{2}}^{\eta_{\rm f}} .
\end{align}
Note that the $+1$ in the eigenvalues $-(n^2-1)$ of the spatial mode functions is exactly canceled by the mass term contribution in the action, which appears due to the spatial curvature of the space $\mathbb{S}^3$. Now, the equation of motion corresponding to each of the modes is
\begin{align} \label{eq:perturbation_eom} 
\phi''_{n\ell m} + n^2 \phi_{n\ell m} = 0~. 
\end{align}
The solution to the above equation of motion is given by the plane waves
\begin{align}
    \e^{\pm i n \eta}.
\end{align}
These solutions only depend on the principal quantum number $n$ and do not explicitly depend on the quantum numbers $(\ell,m)$. Regularity of the solutions at $t=0$, or $\eta\to+i\infty+\eta_{\rm f}/2$ forces the choice of the only valid mode:
\begin{align}
    \e^{+i n \eta},
\end{align}
which is the correct Bunch-Davies mode selected with the standard $i\epsilon$ prescription in the flat FLRW case, as discussed above. Since the solution to the free theory must satisfy the boundary conditions \ref{eq:free-boundary-condition}, the solution for the time dependent mode functions $ \phi_{n\ell m}(\eta)$ are given by 
\begin{align}\label{eq:phi-nlm}
    \phi_{n\ell m}(\eta) = \Phi_{n\ell m} \e^{i n (\eta-\eta_{\rm f})} ,
\end{align}
where $\Phi_{n\ell m}$ are the analogues of Fourier coefficients on $\mathbb{S}^3$ or the Harmonic coefficients of the field distribution at the final hypersurface, defined as follows
\begin{align}\label{eq:boundary-phi}
    \Phi(\boldsymbol{x}) = \sum_{n\ell m} \Phi_{n\ell m}Q_{n\ell m}(\boldsymbol{x}) .
\end{align}
The full solution for the background field is then
\begin{align}
    \phi_{\circ}(x) = \sum_{n\ell m} \Phi_{n\ell m} \e^{i n (\eta-\eta_{\rm f})} Q_{n\ell m}(\boldsymbol{x}) .
\end{align}
Plugging this solution back into the action we get
\begin{align}
    \mathcal{S}_{\rm free}[\Phi] = \frac{i}{2}\sum_{n\ell m} n |\Phi_{n\ell m}|^2 .
\end{align}
However, after reversing the field rescaling in \ref{eq:field-rescaling-1}, \ref{eq:field-rescaling-2}, we get
\begin{subequations}
\begin{align}
    \mathcal{S}_{\circ}^{\rm ESU} = & R_{\mathbb{S}^3}^2\frac{i}{2}\sum_{n\ell m} n |\Phi_{n\ell m}|^2 \\
    \mathcal{S}_{\circ}^{\rm HH} = & q(\eta_{\rm f})\frac{i}{2}\sum_{n\ell m} n |\Phi_{n\ell m}|^2 \label{eq:gaussian-free-theory-scale}
\end{align}   
\end{subequations}
Then, the wave function for the free theory is the following
\begin{align}\label{eq:gaussian-free-theory}
    \Psi_{\circ}[\Phi]=\e^{i \mathcal{S}_{\rm free}[\Phi]} = \begin{cases}
        \prod_{n\ell m} \e^{-\frac{3}{2\Lambda} n |\Phi_{n\ell m}|^2}, & \text{for ESU} \\
        \prod_{n\ell m} \e^{-\frac{q(\eta_{\rm f})}{2} n |\Phi_{n\ell m}|^2}, & \text{for HH}
    \end{cases}  ,
\end{align}
which is a Gaussian distribution over the absolute square of the Harmonic coefficients of the final field distribution $\Phi(\boldsymbol{x})$ for all quantum numbers $(n\ell m)$. Note that the Hartle-Hawking wave function, in the free theory, agrees with that of the Einstein static universe in the limit $q(\eta_{\rm f})\to 3/\Lambda$.

We now proceed to discuss the interacting part of the wave function.

\subsection{Propagators}
\label{sec:propagators}
The free solutions, satisfying the boundary conditions \eqref{eq:free-boundary-condition}, can be expressed as
\begin{align}
    \phi_{\circ}(\eta,\boldsymbol{x}) = & \int_{\mathbb{S}^3} d^3\boldsymbol{y}\sqrt{h} \, K(\boldsymbol{x},\boldsymbol{y},\eta) \Phi(\boldsymbol{y}) ,
\end{align}
where $K(\boldsymbol{x},\boldsymbol{y},\eta)$ is called the \textit{bulk-to-boundary} propagator. For this to be the solution, the bulk-to-boundary propagator must satisfy the following properties
\begin{subequations}
\begin{align}
    \mathcal{O}(\boldsymbol{x},\eta)K(\boldsymbol{x},\boldsymbol{y},\eta) & = 0, \\ 
    \lim_{\eta\to \eta_{\rm f}}K(\boldsymbol{x},\boldsymbol{y},\eta) & = \frac{1}{\sqrt{h}}\delta^{(3)}(\boldsymbol{x}-\boldsymbol{y}), \\
    \lim_{\eta\to +i\infty+\frac{\eta_{\rm f}}{2}}K(\boldsymbol{x},\boldsymbol{y},\eta) & = 0,
\end{align}    
\end{subequations}
where $\mathcal{O}(\boldsymbol{x},\eta)\phi_{\circ}(\boldsymbol{x},\eta)=0$ is the free theory equation of motion. It is convenient to go to the Harmonic space on $\mathbb{S}^3$, where the bulk-to-boundary propagator is expressed as
\begin{align}
    K(\boldsymbol{x},\boldsymbol{y},\eta) = \sum_{n\ell m} K_{n\ell m}(\eta) Q_{n\ell m}(\boldsymbol{x})Q^*_{n\ell m}(\boldsymbol{y}) ,
\end{align}
where \eqref{eq:phi-nlm}, \eqref{eq:boundary-phi} imply that $K_{n\ell m}$ satisfies the boundary conditions
\begin{align}
    \lim_{\eta\to \eta_{\rm f}}K_{n\ell m}(\eta) = 1 , ~~~~ \lim_{\eta\to +i\infty+\frac{\eta_{\rm f}}{2}}K_{n\ell m}(\eta) = 0.
\end{align}
Then the free theory solution in the real space can be rewritten in the Harmonic space as
\begin{align}
    \phi_{\circ}(\eta,\boldsymbol{x}) = & \int d^3\boldsymbol{y}\sqrt{h} \, K(\boldsymbol{x},\boldsymbol{y},\eta) \Phi(\boldsymbol{y}) \nonumber\\
    = & \sum_{n\ell m} K_{n\ell m}(\eta) \Phi_{n\ell m} Q_{n\ell m}(\boldsymbol{x}) ,
\end{align}
where we have used the relation
\begin{align}
    \frac{1}{\sqrt{h}}\delta^{(3)}(\boldsymbol{x}-\boldsymbol{y}) = \sum_{n\ell m} Q_{n\ell m}(\boldsymbol{x}) Q^*_{n\ell m}(\boldsymbol{y}) .
\end{align}
Further, recalling the \textit{ansatz} \ref{phix}, we can write the time-dependent mode functions in terms of the bulk-to-boundary propagator as
\begin{align}
    \phi_{n\ell m}(\eta) = K_{n\ell m}(\eta)\Phi_{n\ell m} ,
\end{align}

Then we can write the bulk-to-boundary propagator for the Einstein static universe as
\begin{align}
    K_{n\ell m}(\eta) = \e^{i n (\eta-\eta_{\rm f})} .
\end{align}
On the other hand, the propagator corresponding to the free part of $\varphi$ field must satisfy the following properties
\begin{subequations}
\begin{align}
    \mathcal{O}(\boldsymbol{x}_{1},\eta_{1})G(\boldsymbol{x}_{1},\eta_{1},\boldsymbol{x}_{2},\eta_{2}) & = \frac{\delta^{(3)}(\boldsymbol{x}_{1}-\boldsymbol{x}_{2})}{\sqrt{h}}\nonumber\\
    & ~~~~\times \delta(\eta_{1}-\eta_{2}), \\
    \lim_{\eta_{1,2}\to +i\infty+\frac{\eta_{\rm f}}{2}, \eta_{\rm f}}G(\boldsymbol{x}_{1},\eta_{1},\boldsymbol{x}_{2},\eta_{2}) & = 0 .
\end{align}    
\end{subequations}
This propagator is known as the \textit{bulk-to-bulk} propagator since the Dirac delta sources live in the bulk, and it vanishes on the initial and final time boundary hypersurfaces. The explicit form of this propagator in the Harmonic space is derived in \ref{app:propagators}.

\subsection{Wave function coefficients}
Since each of the $\phi_{\circ}(\eta_{s},\boldsymbol{x}_{s})$ can be expressed as a sum over the Harmonic coefficients of the final distribution, $\Phi_{n\ell m}$, the wave function can be expressed as
\begin{align}\label{eq:general-wave-function-coefficeints}
    \Psi[\Phi] = & \e^{-\frac{q(\eta_{\rm f})}{2}\sum_{\boldsymbol{{\sf n}}} n |\Phi_{\boldsymbol{{\sf n}}}|^2} \exp\Bigg[\sum_{r\geq 3}\frac{1}{r!}\left(\prod_{i=1}^{r} \sum_{\boldsymbol{{\sf n}}_i}\Phi_{\boldsymbol{{\sf n}}_{i}}\right) \nonumber\\
    &~~~~~~\times  \psi^{(r)}(\{ \boldsymbol{{\sf n}}_{i}\})\Bigg] .
\end{align}
Here we have used the notation $\boldsymbol{{\sf n}}=(n,\ell,m)$, and $\boldsymbol{{\sf n}}_{i}=(n_{i},\ell_{i}, m_{i})$ to denote all three quantum numbers together, and $\{\boldsymbol{{\sf n}}_{i}\}$ contains all the quantum numbers from $i=1$ to $i=r$, that is, $\{\boldsymbol{{\sf n}}_{i}\}=(n_{1},\ell_{1}, m_{1}; n_{2},\ell_{2}, m_{2};\dots;n_{r},\ell_{r}, m_{r})$. The quantities $\psi^{(r)}$ are called the wave function coefficients, and will be central objects of our computation below. Note that since the wave function is expressed in the exponentiated form, the $\psi^{(r)}$ only correspond to the connected diagrams.

For the Einstein static universe case, the interaction term in the action, after splitting into free and fluctuating pieces, as in \eqref{eq:scalar-flucctuations}, has the form
\begin{align}
    \mathcal{S}_{\rm int} = \int_{\mathcal{C}(+i\infty+\frac{\eta_{\rm f}}{2},\eta_{\rm f})}\rd\eta\rd^3\boldsymbol{x}\sqrt{h}\frac{\lambda_{p}}{p!} R^{4-p}_{\mathbb{S}^3} \sum_{q=0}^{p} \begin{pmatrix}
        p\\
        q
    \end{pmatrix} \phi_{\circ}^{p-q} \varphi^{q} .
\end{align}
At each order in $\lambda_{p}$ the contribution to the wave function is
\begin{align}
    & \langle\mathcal{S}_{\rm int}^j\rangle = \mathscr{N} \int_{\varphi(+i\infty+\frac{\eta_{\rm f}}{2})=0}^{\varphi(\eta_{\rm f})=0} \mathcal{D}[\varphi] \e^{i\mathcal{S}_{\rm free}[\varphi]}\prod_{s=1}^{j}\int_{\mathcal{C}(+i\infty+\frac{\eta_{\rm f}}{2},\eta_{\rm f})} \nonumber\\
    &~~\rd\eta_s\rd^3\boldsymbol{x}_s \sqrt{h}\frac{\lambda_{p}}{p!} R^{4-p}_{\mathbb{S}^3} \sum_{q_s=0}^{p} \begin{pmatrix}
        p\\
        q_s
    \end{pmatrix} \phi_{\circ}^{p-q_s}(\eta_{s},\boldsymbol{x}_{s}) \varphi^{q_s}(\eta_{s},\boldsymbol{x}_{s}) .
\end{align}
Substituting the above expression in \eqref{eq:expansion}, we can compute the perturbative expansion order by order. As $\mathcal{S}_{\rm free}[\varphi]$ is a quadratic action, when the integrand is expanded, the only terms with an even total number of fluctuation fields $\varphi$ have a non-zero contribution. For the wave function coefficients defined in \eqref{eq:general-wave-function-coefficeints} (with $q(\eta_{\rm f})=3/\Lambda$ for ESU), we only need the connected diagrams, which we will calculate in the next section.

The computation of the wave function coefficients on the Hartle-Hawking saddle is similar to the Einstein static universe case except for the fact that the coupling factors are time-dependent. This time dependence can be traded in favour of a sum over frequencies due to the relation \ref{eq:spectral}. Then the interaction contribution to the wave function in the Hartle-Hawking saddle can be expressed in a similar way as
\begin{align}
    \langle\mathcal{S}_{\rm int}^j\rangle =&  \mathscr{N} \int_{\varphi(+i\infty+\frac{\eta_{\rm f}}{2})=0}^{\varphi(\eta_{\rm f})=0} \mathcal{D}[\varphi] \e^{i\mathcal{S}_{\rm free}[\varphi]}\prod_{s=1}^{j}\int_{\mathcal{C}(+i\infty+\frac{\eta_{\rm f}}{2},\eta_{\rm f})} \nonumber\\
    & \rd\eta_s\rd^3\boldsymbol{x}_s\sqrt{h}\sum_{k_s=0}^{\infty}\frac{\tilde{\lambda}_{p}(\omega_{k_s})}{p!} \e^{i\omega_{k_s} \eta_s} \sum_{q_s=0}^{p} \begin{pmatrix}
        p\\
        q_s
    \end{pmatrix} \nonumber\\
    &\times \phi_{\circ}^{p-q_s}(\eta_{s},\boldsymbol{x}_{s}) \varphi^{q_s}(\eta_{s},\boldsymbol{x}_{s}) .
\end{align}
Since every factor of the coupling coefficient corresponds to a vertex in the Feynman-Witten diagram, every vertex in a Hartle-Hawking saddle is given by that of an Einstein static universe but with additional sums, and the factor $\e^{i\omega_k \eta}$ shifts the energy associated with that vertex by $\omega_{k}$. Schematically, if the wave function coefficients are expressed as a function of energies at vertices $X_{i}$ with $i=1,\dots, V$, then
\begin{align}
    \psi^{(r)}_{\rm HH} = \left(\prod_{i=1}^{V}\sum_{k_{i}=0}^{\infty}\tilde{\lambda}(\omega_{k_{i}})\e^{i\omega_{k_{i}}\eta_{\rm f}}\right) \psi^{(r)}_{\rm ESU}(\{X_{i}+\omega_{k_i}\}).
\end{align}
Relations of this type are interesting because of the fact that ESU wave function coefficients $\psi^{(r)}_{\rm ESU}$ are rational functions of the external kinematics, which are obtainable via flat-space cosmological polytopes. Then the time-dependence of the Hartle-Hawking saddle maps the dynamics into discrete, infinite spectral sums over rational poles. As we shall show below, these nested sums are the mathematical definition of generalized multi-variable hypergeometric series. In the following, we compute relations of this type and explicitly discuss the nature of these additional sums for the case of a cubic interaction $p=3$.

In the usual case, the wave function coefficient of the FLRW spacetime with flat spatial geometry is expressed as integrals over the shifted kinematics of the flat spacetime wave function coefficients \cite{arkani-hamed_cosmological_2017,arkani-hamed_differential_2024}. However, in the present case, since the mode functions on $\mathbb{S}^3$ have a discrete spectrum, and the wave function coefficients are functions of kinematic variables with discrete values, it is meaningful to replace the continuous integrals with discrete sums.

\section{Wave function coefficients as discrete spectral sums}
\label{sec:sums}
In this section, we first compute the wave function coefficients on the Einstein static universe, and then express the coefficients on the Hartle-Hawking saddles as discrete spectral sums over them, which we identify with multi-variable hypergeometric series.

\subsection{Wave function coefficients in Einstein static universe}
\label{sec:ESU}
The time dependent mode functions $\e^{\pm i n\eta}$ for a conformally coupled scalar field in an Einstein static universe are the same as that of a massless scalar field in the flat spacetime, with the continuous wave number labeling the plane waves in $\mathbb{R}^3$ replaced by the discrete principal quantum numbers for the hyper-spherical harmonics on $\mathbb{S}^3$
\begin{align}
    k \leftrightarrow n .
\end{align}
As a result, the propagators, and hence the wave function coefficients, in the two cases have the same mathematical form up to the replacement $k\to n$.
The interaction terms can be represented with Feynman-Witten diagrams. For example, the diagram with one vertex and three external lines corresponds to the following term
\begin{align}
    \raisebox{-29pt}{
\begin{tikzpicture}[line width=1. pt, scale=2]
\draw[lightgray, line width=1.pt] (0,0) -- (-0.25,0.55);
\draw[lightgray, line width=1.pt] (0,0) -- (0.0,0.55);
\draw[lightgray, line width=1.pt] (0,0) -- (0.25,0.55);
\draw[lightgray, line width=2.pt] (-0.5,0.55) -- (0.5,0.55);
\draw[fill=black] (0,0) circle (.03cm);
\node[scale=1] at (0,-.15) { };
\end{tikzpicture}} = & i \frac{\lambda_{3}}{3!} \int_{\mathcal{C}(+i\infty+\frac{\eta_{\rm f}}{2},\eta_{\rm f})}\rd \eta \int_{\mathbb{S}^3}\rd^3 \boldsymbol{x}\sqrt{h} \,  (\phi_{\circ}^3 ) .
\end{align}
Here the horizontal line represents the final boundary hypersurface at $\eta_{\rm f}$, and the dot represents the interaction vertex inside the bulk.
It is convenient to express the above integral in the Harmonic space as
\begin{align}
     & i \frac{\lambda_{3}}{3!} \int_{\mathcal{C}(+i\infty+\frac{\eta_{\rm f}}{2},\eta_{\rm f})}\rd \eta \int_{\mathbb{S}^3}\rd^3 \boldsymbol{x}\sqrt{h} \,  (\phi_{\circ}^3 ) \nonumber\\
    & = i \frac{\lambda_{3}}{3!} \left(\prod_{i=1}^{3} \sum_{n_{i}\ell_{i} m_{i}}\Phi_{n_{i}\ell_{i} m_{i}}\right) \int_{\mathcal{C}(+i\infty+\frac{\eta_{\rm f}}{2},\eta_{\rm f})}\rd \eta  \nonumber\\
    &~~ \times \e^{i (\eta-\eta_{\rm f})\sum_{i=1}^{3} n_{i} }\int_{\mathbb{S}^3}\rd^3 \boldsymbol{x}\sqrt{h} \,    \prod_{i=1}^{3}Q_{n_{i}\ell_{i} m_{i}}(\boldsymbol{x}) .
\end{align}
Here, each of the external lines contribute a factor of the bulk-to-boundary propagator. Also, we have absorbed the factor $R_{\mathbb{S}^3}$ into the coupling constant. Compare this with the flat spacetime computation for the same diagram
\begin{align}
& i \frac{\lambda_{3}}{3!} \int_{-\infty(1-i\epsilon)}^{0} \rd \eta \int_{\mathbb{R}^3}\rd^3 \boldsymbol{x} \,  (\phi_{\circ}^3 ) \nonumber\\
& = i\frac{\lambda_{3}}{3!} \left(\prod_{i=1}^{3} \int\frac{\rd^3\boldsymbol{k}_{i}}{(2\pi)^3}\Phi(\boldsymbol{k}_{i})\right) \int_{-\infty(1-i\epsilon)}^{0}\rd \eta \, \nonumber\\
&~~~~~~ \times \e^{i \eta \sum_{i=1}^{3} k_{i}} (2\pi)^3 \delta^{(3)}(\sum_{i}^{}\boldsymbol{k}_{i})
\end{align}
Notice that the Dirac delta function which came from the spatial integral over the flat space mode functions $\e^{i\left(\sum\boldsymbol{k}\right)\boldsymbol{\cdot x}}$ preserves the linear momentum conservation at each vertex. In the case of $\mathbb{S}^3$, the Dirac delta function is replaced by the spatial integrals of three hyperspherical harmonics
\begin{align}
    \int_{\mathbb{S}^3}\rd^3 \boldsymbol{x}\sqrt{h} \,    \prod_{i=1}^{3}Q_{n_{i}\ell_{i} m_{i}}(\boldsymbol{x}) .
\end{align}
The above integral is a $\mathbb{S}^3$ generalization of Gaunt coefficients on $\mathbb{S}^2$, and provides angular momentum selection rules corresponding to the symmetry group $SO(4)$. These selection rules, for the integral to be non-vanishing, are
\begin{subequations}\label{eq:angular-momentum-selection}
    \begin{align}
        |n_1 - n_2| \leq n_3 - 1 &  \leq n_1 + n_2 - 2 , \\
        n_1 + n_2 + n_3 - 3 & \in 2\mathbb{Z} , \\
        |\ell_1 - \ell_2| \leq \ell_3 & \leq \ell_1 + \ell_2 , \\
        \ell_1 + \ell_2 + \ell_3 & \in 2\mathbb{Z} , \\
        m_1 + m_2 + m_3 & = 0 \label{eq:spin-conserve}.
    \end{align}
\end{subequations}
For more details on the integrals over the hyperspherical harmonics, among other properties, see, for example, \cite{wen_properties_1985}.

In the literature, it is conventional to strip off from the wave function coefficients $\psi^{(r)}$ the external field distribution coefficients, coupling factor and the delta function and discuss only the time integral part, which encodes the bulk dynamics and depends only on the external and internal kinematical variables. 

For the Hartle-Hawking case as well, we can similarly define the Harmonic space wave function coefficient as:
\begin{center}
 \raisebox{-29pt}{
\begin{tikzpicture}[line width=1. pt, scale=2]
\draw[lightgray, line width=1.pt] (0,0) -- (-0.25,0.55);
\draw[lightgray, line width=1.pt] (0,0) -- (0.0,0.55);
\draw[lightgray, line width=1.pt] (0,0) -- (0.25,0.55);
\draw[lightgray, line width=2.pt] (-0.5,0.55) -- (0.5,0.55);
\draw[fill=black] (0,0) circle (.03cm);
\node[scale=1] at (-0.25,0.65) {$n_{1}$};
\node[scale=1] at (0,0.65) {$n_{2}$};
\node[scale=1] at (0.25,0.65) {$n_{3}$};
\node[scale=1] at (0.0,-0.15) {$X_{1}=n_{1}+n_{2}+n_{3}$};
\end{tikzpicture}}
\end{center}
\begin{align}
    \tilde{\psi}^{(1)}_{\rm ESU} = i\int_{\mathcal{C}(+i\infty+\frac{\eta_{\rm f}}{2},\eta_{\rm f})}\rd \eta \, \e^{i (\eta-\eta_{\rm f})\sum_{i=1}^{3} n_{i} } .
\end{align}
These coefficients can be thought of as the momentum space diagrams. Unlike the superscript $r$ in the wave function coefficients $\psi^{(r)}$, it is useful to represent the reduced coefficients as $\tilde{\psi}^{(V)}_{\rm ESU}(\{X_i\},\{Y_{jk}\})$, where $V$ is the number of vertices in the diagram, and $X_i$ is the sum of principal quantum numbers associated with the external lines at the $i$-th vertex, and $Y_{jk}$ is the principal quantum number associated with the internal line between vertices $j$ and $k$. In the present case, $\tilde{\psi}^{(1)}_{\rm ESU}=\tilde{\psi}^{(1)}_{\rm ESU}(X_1)$, where $X_{1}=n_1+n_2+n_3$.

The next tree diagram with four external fields comes from an integral of the form
\begin{align}
& - \frac{1}{2} \left(\frac{\lambda_{3}}{2}\right)^2 \int_{\mathcal{C}(+i\infty+\frac{\eta_{\rm f}}{2},\eta_{\rm f})}\rd \eta \int_{\mathcal{C}(+i\infty+\frac{\eta_{\rm f}}{2},\eta_{\rm f})}\rd \eta' \int_{\mathbb{S}^3}\rd^3 \boldsymbol{x}\sqrt{h} \nonumber\\
& ~~~~ \int_{\mathbb{S}^3}\rd^3 \boldsymbol{y}\sqrt{h} \,\phi_{\circ}^2(\boldsymbol{x}) \langle \varphi(\boldsymbol{x})\varphi(\boldsymbol{y})\rangle \phi_{\circ}^2(\boldsymbol{y}) .
\end{align}
In the Harmonic space representation, each of the external lines contributes a factor of the bulk-to-boundary propagator $K_{\boldsymbol{{\sf n}}}(\eta)$, and the internal lines contribute factors of bulk-to-bulk propagators $-i G_{\boldsymbol{{\sf n}}}(\eta,\eta')$:
\begin{center}
\scalebox{1.0}{
 \raisebox{-26pt}{
\begin{tikzpicture}[line width=1. pt, scale=2]
\draw[fill=black] (0,0) -- (1,0);
\draw[fill=black] (0,0) -- (1,0);
\draw[lightgray,line width=1.pt] (0,0) -- (-0.25,0.55);
\draw[lightgray,line width=1.pt] (0,0) -- (0.25,0.55);
\draw[lightgray,line width=1.pt] (1,0) -- (0.75,0.55);
\draw[lightgray,line width=1.pt] (1,0) -- (1.25,0.55);
\node[scale=1,above] at (-0.25,0.55) {$n_{1}$};
\node[scale=1,above] at (0.25,0.55) {$n_{2}$};
\node[scale=1,above] at (0.75,0.55) {$n_{3}$};
\node[scale=1,above] at (1.25,0.55) {$n_{4}$};
\draw[lightgray, line width=2.pt] (-0.6,0.55) -- (1.6,0.55);
\draw[fill=black] (0,0) circle (.03cm);
\draw[fill=black] (1,0) circle (.03cm);
\node[scale=1] at (0.55,-.15) {$Y_{12}=n$};
\node[scale=1] at (-0.115,-.15) {$X_{1}$};
\node[scale=1] at (1.15,-.15) {$X_{2}$};
\end{tikzpicture}
}
}
\end{center}
\begin{align}
    & -\frac{1}{2}\left(\frac{\lambda_{3}}{2}\right)^2 \sum_{\boldsymbol{{\sf n}}}\left(\prod_{i=1}^{4}\sum_{\boldsymbol{{\sf n}}_{i}}\Phi_{\boldsymbol{{\sf n}}_{i}}\right) \int\int_{\mathcal{C}(+i\infty+\frac{\eta_{\rm f}}{2},\eta_{\rm f})}\rd \eta \, \rd \eta'  \nonumber\\
    &~~ \times K_{\boldsymbol{{\sf n}}_{1}}(\eta) K_{\boldsymbol{{\sf n}}_{2}}(\eta) (-iG_{\boldsymbol{{\sf n}}}(\eta,\eta'))K_{\boldsymbol{{\sf n}}_{3}}(\eta')K_{\boldsymbol{{\sf n}}_{4}}(\eta')  \nonumber\\
    &~~ \times \left(\int_{S^3}\rd^3 \boldsymbol{x}\sqrt{h} \, Q_{\boldsymbol{{\sf n}}}(\boldsymbol{x})\prod_{i=1}^{2}Q_{\boldsymbol{{\sf n}}_{i}}(\boldsymbol{x})\right) \nonumber\\
    &~~ \times \left(\int_{S^3}\rd^3 \boldsymbol{y}\sqrt{h} \, Q^*_{\boldsymbol{{\sf n}}}(\boldsymbol{y})\prod_{i=3}^{4}Q_{\boldsymbol{{\sf n}}_{i}}(\boldsymbol{y})\right) .
\end{align}
Here we define the external energies associated with vertices as $X_{1}=n_{1}+n_{2}$, and $X_{2}=n_{3}+n_{4}$. Again, stripping off the external field coefficients, angular integrals, and coupling factor we can write the reduced coefficient as a time integral
\begin{align}
    \tilde{\psi}_{\rm ESU}^{(2)} = & -\int\int_{\mathcal{C}(+i\infty+\frac{\eta_{\rm f}}{2},\eta_{\rm f})}\rd \eta \, \rd \eta' \, K_{\boldsymbol{{\sf n}}_{1}}(\eta) K_{\boldsymbol{{\sf n}}_{2}}(\eta)\nonumber\\
    & \times (-iG_{\boldsymbol{{\sf n}}}(\eta,\eta'))K_{\boldsymbol{{\sf n}}_{3}}(\eta')K_{\boldsymbol{{\sf n}}_{4}}(\eta') .
\end{align}

We evaluate these time integrals, together with the one for the three-site diagram, in \ref{app:ESU-integrals}.

Then, written as a function of the energies associated with the vertices and internal lines, the reduced wave function coefficients are
\begin{widetext}

\begin{subequations}
\begin{align}
    \tilde{\psi}_{\rm ESU}^{(1)}(X_1) & = \frac{1}{X_{1}} , \label{eq:one-site-ESU}\\
    \tilde{\psi}_{\rm ESU}^{(2)}(X_1,X_2,Y_{12}) & = \frac{1}{(X_{1}+X_{2})(X_{1}+Y_{12})(X_{2}+Y_{12})} , \label{eq:two-site-ESU}\\
    \tilde{\psi}^{(3)}_{\rm ESU}(X_1,X_2,X_{3},Y_{12},Y_{23}) & = \frac{X_{1}+2X_{2}+X_{3}+Y_{12}+Y_{23}}{(X_1+X_2+X_3)(X_{1}+Y_{12})(X_{3}+Y_{23})(X_{2} + Y_{12} + Y_{23})(X_{1}+X_{2}+Y_{23})(X_{2}+X_{3}+Y_{12})} .
\end{align}    
\end{subequations}
\end{widetext}

Here we have used the definitions: for one-site diagram we have
\begin{align}
    X_{1} = n_{1} + n_{2} + n_{3} .
\end{align}
For two-site diagram
\begin{subequations}\label{eq:two-site-energies}
    \begin{align}
        X_{1} & = n_{1} + n_{2}, \\
        X_{2} & = n_{3} + n_{4}, \\
        Y_{12} & = n .
    \end{align}
\end{subequations}
For three-site diagram (see \ref{app:ESU-integrals})
\begin{subequations}
    \begin{align}
        X_{1} & = n_{1} + n_{2} , \\
        X_{2} & = n_{3}, \\
        X_{3} & = n_{4}+n_{5} \\
        Y_{12} & = n_a ,\\
        Y_{23} & = n_b .
    \end{align}
\end{subequations}

Note that the wave function coefficients on the Einstein static universe do not depend on $\eta_{\rm f}$ or, through it, on $q_{1}$. This makes sense since this is a static universe. However, as we will see below, the Hartle-Hawking coefficients do depend on $\eta_{\rm f}$ or $q_{1}$ due to the time dependence of the coupling. Also, due to the correspondence mentioned earlier between the flat space wave number and the principal quantum number of hyperspherical harmonics, the coefficients on the Einstein static universe are the same as those on a flat spacetime with $k$ replaced by $n$. The computation of flat spacetime coefficients is well known in the polytopes program \cite{arkani-hamed_cosmological_2017}.

Moreover, since the principal quantum numbers associated with external lines are restricted to be natural numbers $n_{i}\in \mathbb{N}$, or $n_{i}\geq 1$, and also due to the triangle inequality, the internal lines also satisfy $n_{a}\geq 1$. The rational functions giving the Einstein static universe wave function coefficients have no singularities in the physically allowable domain of the kinematical variables. Further, the condition \ref{eq:spin-conserve} implies the conservation of `spin' angular momentum. For example, for the two-site diagram at site one we have $m_{1}+m_{2}+m=0$, and at site two we have $-m+m_3+m_4=0$. Together, they enforce $m_{1}+m_{2}+m_{3}+m_{4}=0$.

In the following subsection, we will discuss the sums over these rational functions that reproduce the corresponding coefficients on the Hartle-Hawking saddle.

\subsection{Wave function coefficients on the complex Hartle-Hawking saddles}
\label{sec:HH}
We discussed in \ref{sec:ESU-vs-HH} that a conformally coupled field on a Hartle-Hawking saddle is mathematically equivalent to the same on the Einstein static universe with a time-dependent coupling. This time dependence can be replaced with a sum using the decomposition \ref{eq:spectral}, which reads
\begin{align}
    \lambda_{p}(\eta) = \sum_{k=0}^{\infty} \tilde{\lambda}_{p}(\omega_{k}) \, \e^{i\omega_{k}\eta} .
\end{align}
Since the bulk-to-boundary propagators are normalized at the final hypersurface, it is convenient to split the vertex factor as $\e^{i\omega_{k}\eta} = \e^{i\omega_{k}\eta_{\rm f}}\,\e^{i\omega_{k}(\eta-\eta_{\rm f})}$. The factor $\e^{i\omega_{k}(\eta-\eta_{\rm f})}$ then shifts the external energy associated with a vertex by $+\omega_{k}=2k+1$ in the Einstein static universe, while the residual factor $\e^{i\omega_{k}\eta_{\rm f}}$ gives a $k$-dependent phase in the spectral sum. For example, the one-site diagram on the Hartle-Hawking saddle becomes
\begin{center}
\raisebox{-29pt}{
\begin{tikzpicture}[line width=1.0pt, scale=2]
\shade[top color=green!20, bottom color=red!20]
    (0.5,0.55) arc (0:-180:0.65) -- cycle;
\draw[lightgray, thick, dashed] (0.5,0.55) arc (0:-180:0.65);
\draw[lightgray, line width=1.0pt] (-0.15,0.050) -- (-0.45,0.55);
\draw[lightgray, line width=1.0pt] (-0.15,0.050) -- (-0.15,0.55);
\draw[lightgray, line width=1.0pt] (-0.15,0.050) -- (0.15,0.55);
\draw[lightgray, line width=2.0pt] (-0.81,0.55) -- (0.51,0.55);
\draw[fill=black] (-0.15,0.050) circle (.03cm);
\node[scale=1,above] at (-0.45,0.55) {$n_{1}$};
\node[scale=1,above] at (-0.15,0.55) {$n_{2}$};
\node[scale=1,above] at (0.15,0.55) {$n_{3}$};
\end{tikzpicture}}
\end{center}
\begin{align}
    & \frac{i}{3!} \sum_{k=0}^{\infty} \tilde{\lambda}_{3}(\omega_{k}) \int_{\mathcal{C}(+i\infty+\frac{\eta_{\rm f}}{2},\eta_{\rm f})}\rd \eta \int_{\mathbb{S}^3}\rd^3 \boldsymbol{x}\sqrt{h} \,  \e^{i\omega_{k}\eta}(\phi_{\circ}^3 ) \nonumber\\
    = & \frac{\lambda_{3}^{\rm HH}\sqrt{\frac{3}{\Lambda}}}{3!} \left(\prod_{i=1}^{3} \sum_{n_{i}\ell_{i} m_{i}}\Phi_{n_{i}\ell_{i} m_{i}}\right)\sum_{k=0}^{\infty}\frac{2(-1)^k \e^{i(2k+1)\eta_{\rm f}}}{(2k+1) + \sum_{i=1}^{3} n_{i}} \nonumber\\
    &~~~~\times\int_{\mathbb{S}^3}\rd^3 \boldsymbol{x}\sqrt{h} \,    \prod_{i=1}^{3}Q_{n_{i}\ell_{i} m_{i}}(\boldsymbol{x})
\end{align}
As a result, the Hartle-Hawking coefficients are expressed as a phase-weighted sum over shifts of the external energies, with the summand being the coefficients on the Einstein static universe. For the one-site diagram, we have
\begin{align}\label{eq:one-site-sum-HH}
    \tilde{\psi}^{(1)}_{\rm HH}(X) & = \sum_{k=0}^{\infty} 2 (-1)^k \e^{i(2k+1)\eta_{\rm f}}\, \tilde{\psi}^{(1)}_{\rm ESU}(X+(2k+1)) ,
\end{align}
where $X=n_{1}+n_{2}+n_{3}$ and $\tilde{\psi}^{(1)}_{\rm ESU}(X)$ given by \eqref{eq:one-site-ESU}. The colour gradient in the diagram represents the fact that for the Hartle-Hawking saddle, the conformal time is complex.
Then for the two-site diagram we obtain
\begin{center}
\scalebox{0.85}{
 \raisebox{-26pt}{
\begin{tikzpicture}[line width=1.0pt, scale=2]
\shade[top color=green!20, bottom color=red!20]
    (1.5,0.55) arc (0:-180:1.0) -- cycle;
\draw[lightgray, thick, dashed] (1.5,0.55) arc (0:-180:1.0);
\draw (0,0) -- (1,0);
\draw[lightgray,line width=1.0pt] (0,0) -- (-0.25,0.55);
\draw[lightgray,line width=1.0pt] (0,0) -- (0.25,0.55);
\draw[lightgray,line width=1.0pt] (1,0) -- (0.75,0.55);
\draw[lightgray,line width=1.0pt] (1,0) -- (1.25,0.55);
\node[scale=1,above] at (-0.25,0.55) {$n_{1}$};
\node[scale=1,above] at (0.25,0.55) {$n_{2}$};
\node[scale=1,above] at (0.75,0.55) {$n_{3}$};
\node[scale=1,above] at (1.25,0.55) {$n_{4}$};
\draw[lightgray, line width=2.0pt] (-0.51,0.55) -- (1.51,0.55);
\draw[fill=black] (0,0) circle (.03cm);
\draw[fill=black] (1,0) circle (.03cm);
\node[scale=1] at (0.5,-.15) {$n$};
\end{tikzpicture}
}
}
\end{center}
\begin{align}
    &\tilde{\psi}^{(2)}_{\rm HH}(X_{1},X_{2},Y_{12}) =  \sum_{k_{1}=0}^{\infty}\sum_{k_{2}=0}^{\infty} 2^2 (-1)^{k_{1}+k_{2}} \e^{i(2k_{1}+2k_{2}+2)\eta_{\rm f}} \nonumber\\
    &~~~~ \times \tilde{\psi}^{(2)}_{\rm ESU}(X_{1}+2k_{1}+1,X_{2}+2k_{2}+1, Y_{12}) ,
\end{align}
where $X_{1}$, $X_{2}$, $Y_{12}$ are defined in \eqref{eq:two-site-energies} and $\tilde{\psi}^{(2)}_{\rm ESU}$ is given by \eqref{eq:two-site-ESU}.

Similarly, for other diagrams in general we have
\begin{align}
     & \tilde{\psi}^{(V)}_{\rm HH}(\{X_{i}\},\{Y_{jk}\}) =\left( \prod_{i=1}^{V} \sum_{k_{i}=0}^{\infty} 2(-1)^{k_{i}} \e^{i(2k_{i}+1)\eta_{\rm f}} \right) \nonumber\\
    &~~~~~~ \times \tilde{\psi}^{(V)}_{\rm ESU}(\{X_{i}+2k_{i}+1\},\{Y_{jk}\}) . \label{eq:HH-sum-general}
\end{align}
It is convenient to collect the phases into the single variable
\begin{align}\label{eq:z-variable}
    z \equiv - \e^{2 i \eta_{\rm f}} , ~~~~ \cos^2\eta_{\rm f} = \frac{3}{\Lambda q_{1}} ,
\end{align}
so that the weight of the $k$-th term at each vertex is $2\e^{i\eta_{\rm f}}z^{k}$. Unlike in the Einstein static universe, these coefficients do depend on the final conformal time $\eta_{\rm f}$, or equivalently on the final scale factor $q_{1}$. The special value $\eta_{\rm f}=0$, that is $z=-1$, corresponds to the final hypersurface sitting at the de Sitter throat $q_{1}=3/\Lambda$, and is the case in which all the phases drop out.

Thus, the (reduced) wave function coefficients on the Hartle-Hawking saddle are readily obtained from the coefficients on the Einstein static universe, which are in turn obtained by replacing the energies with principal quantum numbers in the flat spacetime coefficients that are known from the cosmological polytopes program.

\subsection{Representation of the sums as hypergeometric series}
\label{sec:hypergeometric}
The sums appearing in the wave function coefficients for the Hartle-Hawking saddle can be formally identified as hypergeometric series. In this section we give a few examples of such identifications since it may be helpful to study their properties further (see, for example, \cite{chen_multivariate_2025} for the use of multi-variable hypergeometric series in the case of spatially flat de Sitter cosmological correlators). However, the result of this section has not been used in the discussions of the following sections.

For the one-site diagram, the sum can be rewritten as the following
\begin{align}
    \tilde{\psi}^{(1)}_{\rm HH}(X) = \frac{2\e^{i\eta_{\rm f}}}{X+1}\sum_{k=0}^{\infty}\frac{(1)_{k}\left(\frac{X+1}{2}\right)_{k}}{ \left(\frac{X+3}{2}\right)_{k}}\frac{z^k}{k!},~~~ X=\sum_{i=1}^{3} n_{i} ,
\end{align}
where $z=-\e^{2i\eta_{\rm f}}$ was defined in \ref{eq:z-variable}, and we have used the following definition of Pochhammer symbols and identities
\begin{align}
    (a)_{n} & = a (a+1) \dots (a+n-1) = \frac{\Gamma(a+n)}{\Gamma(a)} .
\end{align}
Recalling the definition of the hypergeometric function
\begin{align}
    {}_n F_{m} \left(
\begin{array}{c|c}
a_{1}, \dots  a_{n} & \multirow{2}{*}{$x$} \\
b_{1}, \dots  b_{m} &
\end{array}
\right) = \sum_{k=0}^{\infty} \frac{(a_{1})_{k} \dots (a_{n})_{k}}{(b_{1})_{k} \dots (b_{m})_{k}}\frac{x^k}{k!},
\end{align}
we can write
\begin{align}
    \tilde{\psi}^{(1)}_{\rm HH}(X) = \left(\frac{2\e^{i\eta_{\rm f}}}{X+1}\right) {}_2 F_{1} \left(
\begin{array}{c|c}
1, \frac{X+1}{2} & \multirow{2}{*}{$z$} \\
\frac{X+3}{2} &
\end{array}
\right) .
\end{align}
Note that the prefactor to the hypergeometric function is given by $(2\e^{i\eta_{\rm f}})^1 \tilde{\psi}^{(1)}_{\rm ESU}(X+1)$, where again $\tilde{\psi}^{(1)}_{\rm ESU}(X)$ given by \eqref{eq:one-site-ESU}. For the two-site tree diagram, we can write
\begin{align}
    & \tilde{\psi}^{(2)}_{\rm HH}(X_{1},X_{2},Y_{12}) = (2\e^{i\eta_{\rm f}})^2 \tilde{\psi}^{(2)}_{\rm ESU}(X_{1}+1,X_{2}+1,Y_{12}) \nonumber\\
    &~~~~\times \sum_{k_{1},k_{2}=0}^{\infty} (1)_{k_{1}} (1)_{k_{2}} \frac{\left(\frac{X_{1}+X_{2}+2}{2}\right)_{k_{1}+k_{2}}}{\left(\frac{X_{1}+X_{2}+4}{2}\right)_{k_{1}+k_{2} }} \frac{\left(\frac{X_{1}+Y_{12}+1}{2}\right)_{k_{1}}}{\left(\frac{X_{1}+Y_{12}+3}{2}\right)_{k_{1}}} \nonumber\\
    & ~~~~ \times \frac{ \left(\frac{X_{2}+Y_{12}+1}{2}\right)_{k_{2}}}{\left(\frac{X_{2}+Y_{12}+3}{2}\right)_{k_{2}}}\frac{z^{k_{1}}z^{k_{2}}}{k_{1}!k_{2}!} .
\end{align}
This double sum can be formally identified with the generalized Kampé de Fériet function \cite{sriv} defined as
\begin{align}
& {}^{p+q}F_{r+s}\left(
\begin{array}{c|c}
a_1,\cdots,a_p\colon b_1,b_1{}';\cdots;b_q,b_q{}' & \multirow{2}{*}{$x,y$} \\
c_1,\cdots,c_r\colon d_1,d_1{}';\cdots;d_s,d_s{}' &
\end{array}\right)  \nonumber\\
& = \sum_{m=0}^\infty\sum_{n=0}^\infty
\frac{(a_1)_{m+n}\cdots(a_p)_{m+n}}{(c_1)_{m+n}\cdots(c_r)_{m+n}}\nonumber\\
&~~~~~~~~\times  \frac{(b_1)_m(b_1{}')_n\cdots(b_q)_m(b_q{}')_n}{(d_1)_m(d_1{}')_n\cdots(d_s)_m(d_s{}')_n}\cdot\frac{x^my^n}{m!n!}.
\end{align}
By this definition the (reduced) coefficient for the 2-site tree diagram is given by
\begin{align}
    & \tilde{\psi}^{(2)}_{\rm HH}(X_{1},X_{2},Y_{12}) = (2\e^{i\eta_{\rm f}})^2 \tilde{\psi}^{(2)}_{\rm ESU}(X_{1}+1,X_{2}+1,Y_{12}) \nonumber\\
    & {}^{1+2}F_{1+1}\left(
\begin{array}{c|c}
\frac{X_{1}+X_{2}+2}{2}:1,1;\frac{X_{1}+Y_{12}+1}{2},\frac{X_{2}+Y_{12}+1}{2} & \multirow{2}{*}{$z,z$} \\
\frac{X_{1}+X_{2}+4}{2}:\frac{X_{1}+Y_{12}+3}{2},\frac{X_{2}+Y_{12}+3}{2} &
\end{array}\right) .
\end{align}
Since the rational functions appearing in the Einstein static universe coefficients can always be re-written in terms of Pochhammer symbols using the relation
\begin{align}
    a+k = \frac{a(a+1)_k}{(a)_k},
\end{align}
the Hartle-Hawking coefficients with an arbitrary number of vertices can always be written as
\begin{align}
    & \tilde{\psi}^{(V)}_{\rm HH}(\{X_{i}\},\{Y_{jk}\}) = (2\e^{i\eta_{\rm f}})^V \tilde{\psi}^{(V)}_{\rm ESU}(\{X_{i}+1\},\{Y_{jk}\}) \nonumber\\
    & \times \left(\begin{array}{c}
         \text{linear combinations of}  \\
        \text{generalized hypergeometric series}
    \end{array}\right) ,
\end{align}
where these multiple series can be formally identified with the generalized Lauricella functions, which generalize the Kampé de Fériet functions to the multivariable case, as introduced and systematically studied by Srivastava and Daoust \cite{srivastava_neumann_1969,srivastava_neumann_1987}.

However, note that these multi-variable hypergeometric series with arguments $\{x_{1},\dots x_{n}\}$ are generally known to converge absolutely strictly inside the region (${\rm max}\{|x_1|,|x_2| \dots |x_{n}| \}< 1$), and in certain cases on the boundary $|x_{i}|=1$. In our case all the arguments evaluate to $x_i = z = -\e^{2i\eta_{\rm f}}$. Since on the final hypersurface $\eta_{\rm f}$ is real and $|z| = 1$, the arguments are on the boundary. We will explicitly discuss the convergence of one- and two-site diagrams in \ref{sec:convergence-late-time}.

\section{Difference equations and kinematic flow}
\label{sec:kinematic-flow}
In \cite{arkani-hamed_differential_2024,arkani-hamed_kinematic_2025}, it was shown that for diagrams at any order, a set of (twisted) integrals, $I_{a}$, form a basis of a finite-dimensional vector space, and under the continuous change of the kinematic parameters, satisfy differential equations of the form
\begin{align}
    \rd I_{a} = A_{ab} I_{b},
\end{align}
where the matrix $A_{ab}$ satisfies the consistency condition $\rd A + A\wedge A=0$. Here the differential operator is defined as $\rd = \sum_{I} \rd Z_{I}\frac{\partial}{\partial Z_{I}}$, where $Z_{I}$ is the set of the kinematical variables including the energies at the internal lines. These equations then determine how the wave function coefficients, and in turn the cosmological correlators, depend on the kinematical variables and change with them, i.e. the kinematical flow. In this description then the bulk time evolution, which we do not observe, has been mapped to the kinematical variables on the late-time hypersurface that we can, in principle, observe. In the present case, where the kinematic variables are forced to be discrete, the precise language of exterior differential calculus seems to be missing. However, below we shall show that analogous relations, where we can express the action of a shift operator on the kinematic variables of a finite set of sums (vector) as a transformation of that vector, exist. These relations would be the discrete difference equations describing discrete kinematic flow of the reduced wave function coefficients.

Before discussing the difference equation and the discrete kinematic flow, we will show that the Hartle-Hawking coefficients satisfy recursion relations at all orders with any number of sites. Notice the fact that the external energies in the sum are shifted by $2k+1$, and that each vertex carries in addition the phase $\e^{i(2k+1)\eta_{\rm f}}$. As a result, shifting the external energies by two units, $X\to X+2$, is the same as a $k\to k+1$ shift in the summand with an additional $-\e^{-2i\eta_{\rm f}}$ coming from the shift $(-1)^{k}\e^{i(2k+1)\eta_{\rm f}}\to - \e^{-2i\eta_{\rm f}}(-1)^{k+1}\e^{i(2(k+1)+1)\eta_{\rm f}}$. For example, in the one-site sum given by \eqref{eq:one-site-sum-HH}, after shifting the vertex energy, we get
\begin{align}\label{eq:sum-one-site}
    \tilde{\psi}^{(1)}_{\rm HH}(X+2) = -\e^{-2i\eta_{\rm f}}\sum_{k=0}^{\infty}\frac{(-1)^{k+1}\e^{i(2(k+1)+1)\eta_{\rm f}}}{\frac{X+1}{2}+(k+1)}.
\end{align}
Now we can use the identity
\begin{align}\label{eq:sum-collapse}
    \sum_{k=k_{\rm i}}^{k_{\rm f}} \Delta_{k}f(k) = -f(k_{\rm i}) + f(k_{\rm f}+1) ,
\end{align}
where $\Delta_{k}f(k)=f(k+1)-f(k)$ is the discrete difference operator, to collapse the sum in \eqref{eq:sum-one-site}, and we get from the boundary values of the summand at $k=0$ and $k=\infty$, the following recursion relation
\begin{align}\label{eq:one-site-recursion}
    \e^{2i\eta_{\rm f}}\tilde{\psi}^{(1)}_{\rm HH}(X+2) + \tilde{\psi}^{(1)}_{\rm HH}(X) & = \frac{2\e^{i\eta_{\rm f}}}{X+1} \nonumber\\
    & = 2\e^{i\eta_{\rm f}}\tilde{\psi}^{(1)}_{\rm ESU}(X+1) .
\end{align}
Repeating this procedure for any diagram with $V$-sites we get the following relation
\begin{align}
    & \left[\prod_{i=1}^{V} (\e^{2i\eta_{\rm f}}S^{(2)}_{X_{i}}+1)\right]\tilde{\psi}^{(V)}_{\rm HH}(\{X_{i}\},\{Y_{jk}\}) \nonumber\\
    & ~~~~~~~~ = \left(2\e^{i\eta_{\rm f}}\right)^{V} \left[\prod_{i=1}^{V} S^{(1)}_{X_{i}}\right] \tilde{\psi}^{(V)}_{\rm ESU}(\{X_{i}\}, \{Y_{jk}\}) , \label{eq:recursion-V}
\end{align}
where we have defined the shift operator as $S^{(s)}_{X_{i}}f(X_{i})=f(X_{i}+s)$. Therefore, the Hartle-Hawking wave function coefficients are constrained by the above recursion relations such that when their external kinematics are shifted by the action of the operator $(\e^{2i\eta_{\rm f}}S^{(2)}_{X_{i}}+1)$ at all vertices, then all the sums collapse, and the final result must be a rational function of the kinematics, obtained from shifting the corresponding Einstein static universe coefficients by the operator $S_{X_{i}}^{(1)}$ at all vertices and multiplying by the overall factor $(2\e^{i\eta_{\rm f}})^{V}$. Now, we will discuss the difference equations and discrete kinematic flow.

At $V=1$, the difference equation or kinematic flow is simplest to demonstrate by defining the vector as $\mathcal{I}^{(1)}(X) = \left(\tilde{\psi}^{(1)}_{\rm HH}(X) ~~~ 1\right)^T $. Then, from \ref{eq:one-site-recursion}, it is easy to see that the action of the operator $S_{X}^{(2)}$, which shifts the kinematic variable by adding $2$ units, can be expressed as a transformation of the vector as
\begin{align}\label{eq:flow-one-site}
    S_{X}^{(2)} \mathcal{I}^{(1)}(X) = \mathscr{A}^{(1)}(X) \mathcal{I}^{(1)}(X) ,
\end{align}
where the transformation matrix (reading off from \eqref{eq:one-site-recursion}) is given by
\begin{align}\label{eq:flow-one-site-matrix}
    \mathscr{A}^{(1)}(X) = \begin{pmatrix}
        -\e^{-2i\eta_{\rm f}} & \frac{2\e^{-i\eta_{\rm f}}}{X + 1} \\
        0 & 1
    \end{pmatrix} .
\end{align}
The above matrix transformation describes the local (discrete) flow of the one-site wave function coefficient with respect to the kinematic variable $X=n_{1}+n_{2}+n_{3}$. In principle, if we know the value of $\mathcal{I}^{(1)}(X)$ or $\tilde{\psi}^{(1)}_{\rm HH}(X)$ at any two independent points in the kinematic space, say $X_{0}$, and $X_{0}+1$ (in general, $X_{0}$, and $X_{0}+\text{odd}$), then the values of the coefficients at any other point in the kinematic space can be determined by a finite number of matrix multiplications. Thus the problem of performing infinite sums reduces to performing matrix multiplications, where the matrix elements are local, rational functions in the kinematic space.

At $V=2$, the discrete kinematic equation is slightly more involved than the example at $V=1$. As we show in \ref{sec:kin-flow-two-site-derivation}, the collection of various sums that, when acted upon with the shift operator, $S_{K}^{(2)}$ where $K=\{X_{1},X_{2},Y_{12}\}$, can be spanned by the same collection, i.e., a set of sums that closes under shift, is the following
\begin{align}
    \mathcal{I}^{(2)} = \begin{pmatrix}
        \mathbf{\Psi} \\ \mathbf{G} \\ \mathbf{F} \\ \mathbf{R}
    \end{pmatrix}
    = \begin{pmatrix}
         \tilde{\psi}^{(2)}_{\rm HH} \\
         G \\
         \psi_{1Y} \\ \psi_{2Y} \\ \psi_{12} \\
         1
    \end{pmatrix} .
\end{align}
Here, the quantity $G$ is a product of two one-site coefficients, defined as follows
\begin{align}\label{eq:G-def}
    G \equiv \frac{\e^{-2i\eta_{\rm f}}}{4}\tilde{\psi}^{(1)}_{\rm HH}(X_{1}+Y_{12})\tilde{\psi}^{(1)}_{\rm HH}(X_{2}+Y_{12}) .
\end{align}
Further, we have also defined $\psi_{1Y} \equiv \tilde{\psi}^{(1)}_{\rm HH}(X_1+Y_{12})$, $\psi_{2Y} \equiv \tilde{\psi}^{(1)}_{\rm HH}(X_2+Y_{12})$, and $\psi_{12} \equiv \tilde{\psi}^{(1)}_{\rm HH}(X_1+X_2+1)$, which are one-site coefficients evaluated at specific combinations of external and internal kinematical variables. Then it can be shown that the action of $S^{(2)}_{K}$ on $\mathcal{I}^{(2)}(X)$ can be spanned by elements of $\mathcal{I}^{(2)}(X_{1},X_{2},Y_{12})$ themselves, with the coefficients being rational and local functions of the kinematical variables, giving us the following discrete kinematical flow equation
\begin{align}\label{eq:flow-two-site}
    S_{K}^{(2)} \mathcal{I}^{(2)} = \mathscr{A}^{(2)}_{K} \mathcal{I}^{(2)} ,
\end{align}
where $K \in \{X_1, X_2, Y_{12}\}$. Here, the transformation matrix can be written in the following block-matrix form:
\begin{align}\label{eq:block-matrix}
    \mathscr{A}_{K}^{(2)} = \begin{pmatrix}
        M_{\Psi\Psi}^K & M_{\Psi G}^K & M_{\Psi F}^K & M_{\Psi R}^K \\
        0 & M_{GG}^K & M_{GF}^K & M_{GR}^K \\
        0 & 0 & M_{FF}^K & M_{FR}^K \\
        0 & 0 & 0 & 1
    \end{pmatrix} .
\end{align}
The form of the above matrix depends on which variable is being shifted. The explicit forms of the sub-blocks for the shifts with respect to the external energies $X_{1}$, $X_{2}$, and the internal energy $Y_{12}$ are given in \ref{app:two-site-matrices}. Note that all the diagonal blocks turn out to be diagonal matrices themselves, so that the transformation matrices are upper triangular.

Equation \eqref{eq:flow-two-site} describes the discrete kinematic flow of the two-site wave function coefficient. Since each elementary shift changes one kinematic variable by two, we need two consecutive starting values for each of the three variables to cover the full kinematic lattice. We therefore specify $\mathcal{I}^{(2)}$ at the eight points
$$(X_1^0+\epsilon_1,\,X_2^0+\epsilon_2,\,Y_{12}^0+\epsilon_3),
    \qquad \epsilon_1,\epsilon_2,\epsilon_3\in\{0,1\}.$$
For any target point, exactly one of these starting points differs from it by an even integer in each coordinate. We can then obtain $\tilde{\psi}^{(2)}_{\rm HH}$ by a finite sequence of matrix multiplications, provided the matrices are well defined along the chosen path and invertible for any backward steps. The selection rules in \eqref{eq:angular-momentum-selection} then determine which configurations contribute to the physical diagram. In particular, the total energies $X_1+Y_{12}$ and $X_2+Y_{12}$ at the two cubic vertices must both be odd.

For this to be true, we also need to satisfy the consistency condition $[S^{(2)}_{K},S^{(2)}_{L}]=0$. Since the transformation matrices depend on the kinematic variables, this translates to the following consistency condition on the transformation matrices
\begin{align}\label{eq:consistency}
    \left(S^{(2)}_{L}\mathscr{A}^{(2)}_{K}\right)\mathscr{A}^{(2)}_{L} = \left(S^{(2)}_{K}\mathscr{A}^{(2)}_{L}\right)\mathscr{A}^{(2)}_{K} ,
\end{align}
where the shift operators act on the kinematic variables in the matrix elements. Physically, this condition ensures that the discrete kinematic flow is independent of the path in the discrete kinematic space: starting from a given point, the value of $\mathcal{I}^{(2)}$ at another point reached by a sequence of shifts does not depend on the order in which the shifts are performed. This is the discrete counterpart of the consistency condition $\rd A+A\wedge A=0$ in the flat FLRW case. We have checked that the explicit matrices given in \ref{app:two-site-matrices} satisfy these consistency relations.

In principle, the above procedure can be followed to obtain the discrete kinematic flow for diagrams with any number of sites. Starting from the wave function coefficient, one shifts its kinematic variables and, using the identity \ref{eq:sum-collapse}, the recursion relations and partial fractions, writes the result as a linear combination of other sums, with coefficients that are rational functions of the kinematic variables. One then shifts the kinematic variables of these new sums, and keeps repeating this procedure until no new sums are generated. The resulting set of sums closes under the shifts, and the corresponding transformation matrices give the discrete kinematic flow. For the one- and two-site diagrams, this procedure terminates with the sets of two and six sums discussed above.

\section{Convergence and the late-time limit}
\label{sec:convergence-late-time}

As we mentioned previously, the wave function coefficients on a Hartle-Hawking saddle can be represented as multi-variable hypergeometric series or functions, called the Srivastava-Daoust series. However, the arguments of these functions lie on the boundary of the region where these functions are known to be convergent. This does not immediately mean that these series are divergent, however, and we will explicitly demonstrate the case for the sums in one- and two-site diagrams.

In the limit $|\eta_{\rm f}|<\pi/2$, that is, for the finite values of the final scale factor $q_{1}<\infty$, the one-site sum can be argued to be convergent by the Dirichlet test. The statement of this test is that for a sum of the form $\sum_{n} a_{n} b_{n}$, where $a_{n}$ is a sequence of monotonically decreasing real numbers towards $0$, and $b_{n}$ can be complex numbers with partial sums $\sum_{n=1}^{K} b_{n}$ being bounded, then the sum is convergent. Writing $\upsilon\equiv\e^{-i\eta_{\rm f}}$, $z=-\e^{2i\eta_{\rm f}}$ as in \ref{eq:z-variable}, we can represent the one-site series sum as
\begin{align}
    \tilde{\psi}^{(1)}_{\rm HH}(X) = 2\upsilon^{-1}\sum_{k=0}^{\infty} c_{k}\,z^{k} ,
    \qquad c_{k} \equiv \frac{1}{2k+X+1} .
\end{align}
We can see that the phases have bounded partial sums,
\begin{align}
    \left|\sum_{k=0}^{K}z^{k}\right| = \left|\frac{1-z^{K+1}}{1-z}\right| \leq \frac{2}{|1-z|}
    = \sec\eta_{\rm f} ,
\end{align}
while $c_{k}>c_{k+1}>0$ and $c_{k}\to0$ as $k\to\infty$. Thus, the Dirichlet test is satisfied for $|\eta_{\rm f}|<\pi/2$. However, in the late-time limit $|\eta_{\rm f}|\to\pi/2$ as we approach the conformal boundary, we see that the partial sum is not bounded as $\sec \eta_{\rm f}$ diverges and the Dirichlet test fails. This seems to be problematic as the late-time limit is a physically relevant one for observations. However, this divergence, as we will see, is purely imaginary and since the cosmological correlators only depend on the real part of the wave function coefficients (see \ref{sec:correlators}), this divergence does not affect the physical result.

It is better to see the limit of the one-site coefficient as we approach the late-time conformal boundary by defining $\eta_{\rm f}=\pi/2-\epsilon$, and in the limit $\epsilon\to 0$ we have
\begin{align}\label{eq:one-site-late-asym}
    \tilde{\psi}^{(1)}_{\rm HH}(X)\simeq \frac{\pi}{2}-\frac{i}{2} \left(2 H_{\frac{X-1}{2}}+2 \log (\epsilon )
   +\log (4)\right) .
\end{align}
Here $H_n$ is the Harmonic number, which for non-integer $n$ is defined as $H_{n}=\sum_{k=1}^{\infty}\left[\frac{1}{k}-\frac{1}{k+n}\right]$. Thus, the logarithmic divergence is purely imaginary, and the physically relevant real part is given by
\begin{align}\label{eq:late-time}
    \lim_{\eta_{\rm f}\to\pi/2} {\rm Re}(\tilde{\psi}^{(1)}_{\rm HH}(X)) = \frac{\pi}{2}
    \qquad\text{for every }X ,
\end{align}
that is, on the late-time conformal boundary the real part of the one-site wave function coefficient becomes independent of the total external energies.

This fact can also be argued from the perspective of the kinematical flow equation derived previously. To determine the kinematic flow of ${\rm Re}(\tilde{\psi}^{(1)}_{\rm HH}(X))$ we take the real part of equation \ref{eq:flow-one-site}, and observe that in the limit $|\eta_{\rm f}|\to \pi/2$, the real part of the transformation matrix ${\rm Re}(\mathscr{A}^{(1)}(X))$ is simply the identity matrix.

For multi-site diagrams, $V\geq2$, as the factors in the denominator grow compared to the numerator due to total energy factors, the sums are expected to be convergent for example for the two-site diagram, we see this by evaluating the corresponding continuous integral
\begin{align}
    &\int_{0}^{\infty}\int_{0}^{\infty} \frac{\rd x \rd y}{(X_{2}+Y_{12}+2y+1)(X_{1}+Y_{12}+2x+1)}\nonumber\\
    &~~~~ \times \frac{1}{(X_{1}+X_{2}+2(x+y)+2)} \nonumber\\
    & = \int_{0}^{\infty} \frac{\rd x [\log(X_{1}+X_{2}+2x+2) -\log(X_{2}+Y_{12}+1)] }{2[(X_{1}+2x+1)^2-Y_{12}^2]} .
\end{align}
Since the integrand for large $x$ goes as $\frac{\log[x]}{x^2}$ and $1/x^2$, the above integral converges and so does the sum for the two-site diagram. Similar argument holds for the higher-site diagrams, the additional total energy factor in the denominator helping with the convergence.

Finally, let us discuss the late-time limit in more detail, as it is the one of most interest for the cosmological correlators. At the late-time conformal boundary $\eta_{\rm f}\to\pm\pi/2$, we have $\upsilon\to\mp i$ and $z\to1$, and the phases in the vertex factors become $(-1)^{k}\e^{i(2k+1)\eta_{\rm f}}\to\pm i$ for every $k$. Therefore, all the shifted coefficients on the Einstein static universe enter the sums with the same weight, and whenever the sums converge at the late-time conformal boundary, as for the two-site diagram, we have
\begin{align}\label{eq:late-time-sum}
     & \tilde{\psi}^{(V)}_{\rm HH}(\{X_{i}\},\{Y_{jk}\}) \nonumber\\
    &~~~~ \to (\pm2i)^{V}\sum_{k_{1},\dots,k_{V}=0}^{\infty} \tilde{\psi}^{(V)}_{\rm ESU}(\{X_{i}+2k_{i}+1\},\{Y_{jk}\}) .
\end{align}
This is the discrete counterpart of the integral representation of the wave function coefficients of a conformally coupled field in de Sitter spacetime with flat slicing, where the flat spacetime coefficients with shifted energies are integrated with a uniform weight \cite{arkani-hamed_cosmological_2017,arkani-hamed_differential_2024}, and for large energies, where the curvature of the spatial slices becomes negligible, the sums with steps of two units indeed reduce to such integrals. In particular, the two-site coefficient becomes real and finite at the late-time conformal boundary,
\begin{align}\label{eq:psi2-late}
    & \tilde{\psi}^{(2)}_{\rm late}(X_{1},X_{2},Y_{12}) \equiv \lim_{\eta_{\rm f}\to\pm\pi/2}\tilde{\psi}^{(2)}_{\rm HH}(X_{1},X_{2},Y_{12}) \nonumber\\
    & ~~ = -4\sum_{k_{1},k_{2}=0}^{\infty}\tilde{\psi}^{(2)}_{\rm ESU}(X_{1}+2k_{1}+1,X_{2}+2k_{2}+1,Y_{12}) ,
\end{align}
and is the same on both the saddles, while for the one-site diagram the sum diverges logarithmically, which is the purely imaginary divergence in \ref{eq:one-site-late-asym}.

Since a shift of the external energies by two units now simply shifts the summation variables without producing any phase, the transformation matrices also simplify at the late-time conformal boundary. For the one-site diagram, the transformation matrix \ref{eq:flow-one-site-matrix} becomes
\begin{align}\label{eq:one-site-late}
    \mathscr{A}^{(1)}(X) \to \begin{pmatrix}
        1 & \mp\frac{2i}{X+1} \\
        0 & 1
    \end{pmatrix} ,
\end{align}
whose real part is the identity matrix, as mentioned above. For the two-site diagram, at the late-time conformal boundary we have $\upsilon^{2}\to-1$, so that $M^{Y_{12}}_{\Psi G}$ given in \ref{app:two-site-matrices} vanishes, and the one-site coefficients enter the shifts of $\tilde{\psi}^{(2)}_{\rm HH}$ only through the combinations
\begin{align}\label{eq:D-def}
    D_{1}\equiv\psi_{1Y}-\psi_{12} , ~~~~ D_{2}\equiv\psi_{2Y}-\psi_{12} ,
\end{align}
whose shifts in turn only involve $D_{1}$, $D_{2}$ and constants. Therefore, at the late-time conformal boundary, the smaller set $\mathcal{I}^{(2)}_{\rm late}=(\tilde{\psi}^{(2)}_{\rm late}, D_{1}, D_{2}, 1)^{T}$ closes under the shifts, and for the $N_{+}$ saddle the transformation matrices acting on it become
\begin{widetext}
\begin{subequations}\label{eq:two-site-late}
\begin{align}
    \mathscr{A}^{(2)}_{X_{1}} & \to \begin{pmatrix}
        1 & 0 & -i c_{X_{1}} & 0 \\
        0 & 1 & 0 & -i (s_{1}-t_{12}) \\
        0 & 0 & 1 & i t_{12} \\
        0 & 0 & 0 & 1
    \end{pmatrix} , \\
    \mathscr{A}^{(2)}_{X_{2}} & \to \begin{pmatrix}
        1 & -i c_{X_{2}} & 0 & 0 \\
        0 & 1 & 0 & i t_{12} \\
        0 & 0 & 1 & -i (s_{2}-t_{12}) \\
        0 & 0 & 0 & 1
    \end{pmatrix} , \\
    \mathscr{A}^{(2)}_{Y_{12}} & \to \begin{pmatrix}
        1-c_{Y} & i c_{Y}(b_{2}+s_{2}/4) & i c_{Y}(b_{1}+s_{1}/4) & c_{Y}(b_{2}s_{1}+b_{1}s_{2}+s_{1}s_{2}/4) \\
        0 & 1 & 0 & -i s_{1} \\
        0 & 0 & 1 & -i s_{2} \\
        0 & 0 & 0 & 1
    \end{pmatrix} ,
\end{align}
\end{subequations}
\end{widetext}
where the coefficients are defined in \ref{app:two-site-matrices}. For the $N_{-}$ saddle, the transformation matrices at the late-time conformal boundary are the complex conjugates of the above. Further, even though the one-site coefficients $\psi_{1Y}$, $\psi_{2Y}$, $\psi_{12}$ diverge logarithmically at the late-time conformal boundary, $D_{1}$ and $D_{2}$ are finite.

In particular, the first rows of $\mathscr{A}^{(2)}_{X_{1}}$ and $\mathscr{A}^{(2)}_{X_{2}}$ in \ref{eq:two-site-late} give an explicit difference equation for the late-time two-site coefficient. Since the logarithmic divergences of the one-site coefficients in \ref{eq:one-site-late-asym} do not depend on the kinematic variables, they cancel in $\psi_{2Y}-\psi_{12}$, and we obtain
\begin{align}\label{eq:late-time-psi2}
    & \tilde{\psi}^{(2)}_{\rm late}(X_{1}+2,X_{2},Y_{12}) - \tilde{\psi}^{(2)}_{\rm late}(X_{1},X_{2},Y_{12}) \nonumber\\
    & ~~~~ = c_{X_{1}}\left[H_{\frac{X_{1}+X_{2}}{2}} - H_{\frac{X_{2}+Y_{12}-1}{2}}\right] ,
\end{align}
and similarly for the shift of $X_{2}$, with the labels $1$ and $2$ exchanged, where $c_{X_{1}}=2/((X_{1}+1)^{2}-Y_{12}^{2})$. Note that, unlike for the one-site diagram, the shift of the real two-site coefficient is sourced by the imaginary parts of the one-site coefficients, through the imaginary elements of the transformation matrices. We give a direct proof of \ref{eq:late-time-psi2}, starting from \ref{eq:psi2-late}, in \ref{app:late-time}. Moreover, since $\tilde{\psi}^{(2)}_{\rm late}$ vanishes for $X_{1}\to\infty$, the difference equation \ref{eq:late-time-psi2} determines the late-time two-site coefficient as a single sum over harmonic numbers,
\begin{align}\label{eq:late-time-single-sum}
    \tilde{\psi}^{(2)}_{\rm late}(X_{1},X_{2},Y_{12}) = & -\sum_{j=0}^{\infty} c_{X_{1}+2j} \nonumber\\
    & \times \left[H_{\frac{X_{1}+X_{2}}{2}+j} - H_{\frac{X_{2}+Y_{12}-1}{2}}\right] ,
\end{align}
which, through \ref{eq:four-point}, gives the four-point function at the late-time conformal boundary.

\section{Cosmological correlators}
\label{sec:correlators}
So far we have discussed the computation of wave function coefficients. Now we show how these coefficients are related to the cosmological correlators \cite{maldacena_non-gaussian_2003,anninos_late-time_2015,stefanyszyn_there_2024,donath_inout_2024,palma_wavefunction_2026} or the $N$-point correlation functions defined as
\begin{align}\label{eq:corr-def}
    \left\langle \Phi_{{\sf a}_{1}}\cdots\Phi_{{\sf a}_{N}}\right\rangle = \frac{\displaystyle\int\mathcal{D}[\Phi]\ \Phi_{{\sf a}_{1}}\cdots\Phi_{{\sf a}_{N}}\ \left|\Psi[\Phi]\right|^{2}}{\displaystyle\int\mathcal{D}[\Phi]\ \left|\Psi[\Phi]\right|^{2}} ,
\end{align}
with the collective index ${\sf a}\equiv(n\ell m)$.

Before proceeding to compute this quantity, we need the reality condition for the scalar field on $\mathbb{S}^{3}$. Let $\bar{\sf a}$ denote the conjugate label, defined by $Q_{\bar{\sf a}}\equiv  Q^{*}_{{\sf a}}$, where $\bar{\sf a}=(n,\ell,-m)$ and we have absorbed the phase $(-1)^{m}$ into the definition. Then we also have $\bar{\bar{\sf a}}={\sf a}$. Reality of $\Phi(\boldsymbol{x})$ then gives
\begin{align}\label{eq:reality}
    \Phi^{*}_{{\sf a}} = \Phi_{\bar{{\sf a}}} ,
\end{align}
which is the analogue of the flat space relation $\Phi^{*}(\boldsymbol{k})=\Phi(-\boldsymbol{k})$. With the generalised Gaunt coefficients
\begin{align}
    \mathcal{G}_{{\sf a}_{1}\cdots{\sf a}_{r}} \equiv \int_{\mathbb{S}^{3}}\rd^{3}\boldsymbol{x}\sqrt{h}\ Q_{{\sf a}_{1}}(\boldsymbol{x})\cdots Q_{{\sf a}_{r}}(\boldsymbol{x}) ,
\end{align}
one has
\begin{align}\label{eq:gaunt-conj}
    \mathcal{G}^{*}_{{\sf a}_{1}\cdots{\sf a}_{r}}=\mathcal{G}_{\bar{\sf a}_{1}\cdots\bar{\sf a}_{r}} ,
    \qquad
    \mathcal{G}^{*}_{\bar{\sf a}_{1}\cdots\bar{\sf a}_{r}}=\mathcal{G}_{{\sf a}_{1}\cdots{\sf a}_{r}} ,
\end{align}
and $\sum_{\sf a}Q_{\sf a}(\boldsymbol{x})Q_{\bar{\sf a}}(\boldsymbol{y})=\delta^{(3)}(\boldsymbol{x}-\boldsymbol{y})/\sqrt{h}$.

Recall that the wave function(al) of the universe expanded in terms of the wave function coefficients has the form
\begin{align}\label{eq:psi-exp}
    \Psi[\Phi] = \exp\left[\sum_{r\geq2}\frac{1}{r!}\left(\prod_{i=1}^{r}\sum_{{\sf a}_{i}}\Phi_{{\sf a}_{i}}\right)\psi^{(r)}({\sf a}_{1},\dots,{\sf a}_{r})\right] ,
\end{align}
with $r=2$ the Gaussian term. Relabelling ${\sf a}_{i}\to\bar{\sf a}_{i}$ in the complex conjugate by \ref{eq:reality}, the modulus square of the wave function can be expressed as
\begin{align}
    \left|\Psi[\Phi]\right|^{2} = \exp\left[\sum_{r\geq2}\frac{1}{r!}\left(\prod_{i=1}^{r}\sum_{{\sf a}_{i}}\Phi_{{\sf a}_{i}}\right)\mathcal{P}^{(r)}({\sf a}_{1},\dots,{\sf a}_{r})\right] ,
\end{align}
where the coefficients $\mathcal{P}^{(r)}$ in the modulus square are
\begin{align}\label{eq:P-def}
    \mathcal{P}^{(r)}({\sf a}_{1},\dots,{\sf a}_{r}) \equiv \psi^{(r)}({\sf a}_{1},\dots,{\sf a}_{r}) + \left[\psi^{(r)}(\bar{\sf a}_{1},\dots,\bar{\sf a}_{r})\right]^{*} .
\end{align}
Recalling the Gaussian wave function (given by the free theory) in \ref{eq:gaussian-free-theory-scale}, we can write
\begin{align}
    \Psi_{\circ}[\Phi] = \exp\left[\frac{1}{2!} \sum_{{\sf a}_{1}} \sum_{{\sf a}_{2}} \Phi_{{\sf a}_{1}}\Phi_{{\sf a}_{2}} \psi^{(2)}({\sf a}_{1},{\sf a}_{2})\right] ,
\end{align}
where we must have
\begin{align}
    \psi^{(2)}({\sf a}_{1},{\sf a}_{2}) = -\frac{1}{2} q_{\rm f}\,n_{1}\,(\delta_{{\sf a}_{1}\bar{\sf a}_{2}}+\delta_{{\sf a}_{2}\bar{\sf a}_{1}}) ,
\end{align}
with $q_{\rm f}\equiv q(\eta_{\rm f})=3/(\Lambda\cos^{2}\eta_{\rm f})$. Then we can write the coefficient in the modulus square of the wave function as
\begin{align}
    \mathcal{P}^{(2)}({\sf a}_{1},{\sf a}_{2}) = & -q_{\rm f}\,n_{1}\,\,(\delta_{{\sf a}_{1}\bar{\sf a}_{2}}+\delta_{{\sf a}_{2}\bar{\sf a}_{1}}) , \\
    \left|\Psi_{\circ}\right|^{2} = & \exp\left[-q_{\rm f}\sum_{\sf a}n_{\sf a}\Phi_{\sf a}\Phi_{\bar{\sf a}}\right] .
\end{align}
Given these, the two-point correlation functions are given after Gaussian integrals as follows
\begin{align}\label{eq:two-point}
    \left\langle \Phi_{{\sf a}}\Phi_{{\sf b}}\right\rangle_{\circ} = P_{{\sf a}}\,\delta_{{\sf a}\bar{{\sf b}}} ,
    \qquad P_{{\sf a}} \equiv \frac{1}{2q_{\rm f}\,n_{{\sf a}}} = \frac{\Lambda\cos^{2}\eta_{\rm f}}{6\,n_{{\sf a}}} .
\end{align}
Now, to compute the higher correlation functions, we can use the Wick's theorem: odd moments vanish and for any even moment
\begin{align}\label{eq:wick}
    \left\langle \Phi_{{\sf a}_{1}}\cdots\Phi_{{\sf a}_{2N}}\right\rangle_{\circ}
    = \sum_{\text{pairings}}\ \prod_{({\sf a}_{i},{\sf a}_{j})} P_{{\sf a}_{i}}\,\delta_{{\sf a}_{i}\bar{{\sf a}}_{j}} ,
\end{align}
where the sum runs over all the $(2N-1)!!$ pairings or contractions. The contraction pairs a label with its conjugate, $\delta_{{\sf a}\bar{\sf b}}$ and not $\delta_{{\sf a}{\sf b}}$, this is the closed-universe
counterpart of $\delta^{(3)}(\boldsymbol{k}_{1}+\boldsymbol{k}_{2})$.

Matching the one-site diagram to \ref{eq:psi-exp}, we can write
\begin{align}\label{eq:psi3}
    \psi^{(3)}({\sf a}_{1},{\sf a}_{2},{\sf a}_{3}) = \lambda_{3}\,\mathcal{C}_{3}\ \tilde{\psi}^{(1)}_{\rm HH}(X)\ \mathcal{G}_{{\sf a}_{1}{\sf a}_{2}{\sf a}_{3}} ,
\end{align}
with $\mathcal{C}_{3}=q_{\rm f}^{3/2}\sqrt{3/\Lambda}$ real and positive. Since $X$ is invariant under ${\sf a}_{i}\to\bar{\sf a}_{i}$, \ref{eq:gaunt-conj} and \ref{eq:P-def} give
\begin{align}\label{eq:P3}
    \mathcal{P}^{(3)}({\sf a}_{1},{\sf a}_{2},{\sf a}_{3}) = \lambda_{3}\,\mathcal{C}_{3}\ \mathcal{G}_{{\sf a}_{1}{\sf a}_{2}{\sf a}_{3}}\ \left[2\,{\rm Re}\,\tilde{\psi}^{(1)}_{\rm HH}(X)\right] .
\end{align}
Expanding the measure to first order and using Wick's theorem \ref{eq:wick}, we can write the 3-point correlator as
\begin{align}
    \left\langle \Phi_{{\sf a}_{1}}\Phi_{{\sf a}_{2}}\Phi_{{\sf a}_{3}}\right\rangle
    &= \left\langle \Phi_{{\sf a}_{1}}\Phi_{{\sf a}_{2}}\Phi_{{\sf a}_{3}} \frac{1}{3!} \sum_{\{{\sf b}\}}\mathcal{P}^{(3)}_{{\sf b}_{1}{\sf b}_{2}{\sf b}_{3}}\Phi_{{\sf b}_{1}}\Phi_{{\sf b}_{2}}\Phi_{{\sf b}_{3}}\right\rangle_{\circ} \nonumber\\
    &= \mathcal{P}^{(3)}(\bar{\sf a}_{1},\bar{\sf a}_{2},\bar{\sf a}_{3})\ \prod_{i=1}^{3}\frac{1}{2q_{\rm f}n_{i}} ,
\end{align}
Putting everything together, in terms of the wave function coefficients the 3-point correlation function becomes
\begin{align}\label{eq:three-point}
    \left\langle \Phi_{{\sf a}_{1}}\Phi_{{\sf a}_{2}}\Phi_{{\sf a}_{3}}\right\rangle = \frac{\lambda_{3}\,\mathcal{C}_{3}\ \mathcal{G}_{\bar{\sf a}_{1}\bar{\sf a}_{2}\bar{\sf a}_{3}}}{\left(2q_{\rm f}\right)^{3}n_{1}n_{2}n_{3}}\ \left[2\,{\rm Re}\,\tilde{\psi}^{(1)}_{\rm HH}(X)\right] .
\end{align}
Similarly, for the four-point function, at the order $\lambda_{3}^2$, two terms contribute
\begin{align}
    \left\langle \Phi_{{\sf a}_{1}}\cdots\Phi_{{\sf a}_{4}}\right\rangle_{\rm c} = \left\langle \Phi_{{\sf a}_{1}}\cdots\Phi_{{\sf a}_{4}}\ \mathcal{V}_{4}\right\rangle_{\circ} + \frac{1}{2!}\left\langle \Phi_{{\sf a}_{1}}\cdots\Phi_{{\sf a}_{4}}\ \mathcal{V}_{3}^{2}\right\rangle_{\circ,\,{\rm c}} ,
\end{align}
where we have written $\mathcal{V}_{r}\equiv\frac{1}{r!}\left(\prod_{i}\sum_{{\sf b}_{i}}\Phi_{{\sf b}_{i}}\right)\mathcal{P}^{(r)}$ and the subscript ${\rm c}$ corresponds to connected diagrams. The result is
\begin{align}\label{eq:four-point}
     &\left\langle \Phi_{{\sf a}_{1}}\cdots\Phi_{{\sf a}_{4}}\right\rangle_{\rm c}= \left(\prod_{i=1}^{4}P_{{\sf a}_{i}}\right)\Bigg[\mathcal{P}^{(4)}(\bar{\sf a}_{1},\bar{\sf a}_{2},\bar{\sf a}_{3},\bar{\sf a}_{4}) \nonumber\\
    &~~~~ + \sum_{(ij|kl)}\ \sum_{{\sf c}}P_{{\sf c}}\ \mathcal{P}^{(3)}(\bar{\sf a}_{i},\bar{\sf a}_{j},{\sf c})\ \mathcal{P}^{(3)}(\bar{\sf a}_{k},\bar{\sf a}_{l},\bar{\sf c})\Bigg] .
\end{align}
Here the symbol $(ij|kl)$ denotes an exchange channel: $(ij)$ at one vertex and $(kl)$ at the other. There are three possible exchanges,
\begin{align}
    (12|34) , \qquad (13|24) , \qquad (14|23) ,
\end{align}
with $(ij|kl)$ and $(kl|ij)$ being the same channel.

Matching the two-site diagram reduced coefficients to the definition \ref{eq:psi-exp}, we can write
\begin{align}\label{eq:psi4}
    \psi^{(4)}({\sf a}_{1},\dots,{\sf a}_{4}) = &\lambda_{3}^{2}\,\mathcal{C}_{4}\sum_{(ij|kl)} \sum_{{\sf c}}\tilde{\psi}^{(2)}_{\rm HH}\!\left(X_{ij},X_{kl},n_{{\sf c}}\right)  \nonumber\\
    &~~~~~~  \times\mathcal{G}_{{\sf a}_{i}{\sf a}_{j}{\sf c}} \mathcal{G}_{{\sf a}_{k}{\sf a}_{l}\bar{\sf c}} ,
\end{align}
with $X_{ij}=n_{i}+n_{j}$ and $\mathcal{C}_{4}=q_{\rm f}^{2}\,(3/\Lambda)$. Then we also have
\begin{align}
    \mathcal{P}^{(4)}({\sf a}_{1},\dots,{\sf a}_{4}) =& \lambda_{3}^{2}\,\mathcal{C}_{4}\sum_{(ij|kl)}\ \sum_{{\sf c}}\left[2\,{\rm Re}\,\tilde{\psi}^{(2)}_{\rm HH}\!\left(X_{ij},X_{kl},n_{{\sf c}}\right)\right] \nonumber\\
    &~~~~~~ \times\mathcal{G}_{{\sf a}_{i}{\sf a}_{j}{\sf c}} \mathcal{G}_{{\sf a}_{k}{\sf a}_{l}\bar{\sf c}} .
\end{align}
Notice that unlike the spatially flat case, as there is no Dirac delta function, the sum over ${\sf c}$ remains. However, this is a finite sum, since the two Gaunt factors in \ref{eq:psi4} require angular momentum selection rules \ref{eq:angular-momentum-selection} to hold.

Further, notice that so far we have kept the final conformal time to be generic $\eta_{\rm f}$. However depending on the saddles, $N_{\pm}$, the final conformal time are related as $\eta_{+}=-\eta_{-}$. Since all the wave function coefficients carry phases of the form $\e^{i(2k+1)\eta_{\rm f}}$, we must have the following relation between the wave function coefficients on different saddles
\begin{align}\label{eq:saddle-conj}
    \tilde{\psi}^{(V)}_{\rm HH}\Big|_{N_{-}} = \left[\tilde{\psi}^{(V)}_{\rm HH}\Big|_{N_{+}}\right]^{*} .
\end{align}
Since the cosmological correlators only depend on the real part of the wave function coefficients, the correlators on the different Hartle-Hawking saddles are the same.

Stripping the Gaunt coefficient, in analogy with stripping the momentum-conserving delta function in flat space, we define the power spectrum and bispectrum as
\begin{align}
    \left\langle \Phi_{{\sf a}_{1}}\Phi_{{\sf a}_{2}}\right\rangle & \equiv P(n_{1})\,\delta_{{\sf a}_{1}\bar{\sf a}_{2}} ,
    \\
    \left\langle \Phi_{{\sf a}_{1}}\Phi_{{\sf a}_{2}}\Phi_{{\sf a}_{3}}\right\rangle & \equiv \mathcal{G}_{\bar{\sf a}_{1}\bar{\sf a}_{2}\bar{\sf a}_{3}}\, B(n_{1},n_{2},n_{3}) ,
\end{align}
Then we have
\begin{align}
    P(n) = \frac{1}{2q_{\rm f}n} , \qquad
    B = \frac{\lambda_{3}\mathcal{C}_{3}}{(2q_{\rm f})^{3}n_{1}n_{2}n_{3}}\left[2\,{\rm Re}\,\tilde{\psi}^{(1)}_{\rm HH}(X)\right] .
\end{align}
Thus, apart from the factor $1/(n_{1}n_{2}n_{3})$, the bispectrum depends on the external quantum numbers \emph{only} through the total energy $X=n_1+n_2+n_3$. In the two limits of interest we have
\begin{align}\label{eq:bispectrum-limits}
    2\,{\rm Re}\,\tilde{\psi}^{(1)}_{\rm HH}(X) \longrightarrow
    \begin{cases}
        H_{\frac{X-1}{4}}-H_{\frac{X-3}{4}} , & \eta_{\rm f}\to0 , \\
        \pi , & \eta_{\rm f}\to\pm\pi/2 .
    \end{cases}
\end{align}
At the de Sitter throat, $\eta_{\rm f}=0$ or $q_{1}=3/\Lambda$, all the phases drop out, $z=-1$, and the one-site sum becomes an alternating series, which gives the first line, while the second line, at the late-time conformal boundary $q_{1}\to\infty$, follows from \ref{eq:late-time}. For large total energies, the first line behaves as $2/X-2/X^{3}+\mathcal{O}(X^{-5})$. Therefore, apart from the factor $1/(n_{1}n_{2}n_{3})$, the bispectrum falls off as $X^{-1}$ for large total energies at the de Sitter throat, while at the late-time conformal boundary its dependence on the total energy drops out.

Since we consider a conformally coupled test field rather than the curvature perturbation, the bispectrum above should not be translated into the usual non-linearity parameter. Nevertheless, its size relative to the power spectra, $B/[P(n_1)P(n_2)+{\rm perms}]$, depends on the external quantum numbers only through the total energy $X$, and falls off as $X^{-1}$ at the late-time conformal boundary.

Similarly, at the late-time conformal boundary, the two-site coefficient entering the four-point function \ref{eq:four-point} is real, and its dependence on the external energies is given by the difference equation \ref{eq:late-time-psi2}, or explicitly by \ref{eq:late-time-single-sum}.

\section{Conclusion}
\label{sec:conclusion}
In this paper, we studied the cosmological correlators of a conformally coupled scalar test field with a cubic interaction on the complex Hartle-Hawking saddles, which dominate the Lorentzian gravitational path integral over homogeneous and isotropic metrics with positive cosmological constant for the initial Neumann and final Dirichlet boundary conditions. On these saddles, the conformal time runs along a complex contour starting at $+i\infty+\eta_{\rm f}/2$, and the contour automatically selects the Bunch-Davies mode functions without needing any $i\epsilon$ prescription.

We showed that a conformally coupled field on the Hartle-Hawking saddle is equivalent to the same field on the Einstein static universe, but with a time-dependent coupling. Since the spatial geometry is $\mathbb{S}^3$, the mode functions in the free theory are plane waves labelled by the discrete principal quantum numbers, and the time dependence of the coupling can be expressed as a discrete sum over frequencies, which is an infinite sum for the cubic interaction. The quartic interaction is conformally invariant and contains no sum, and the higher-order interactions give finite sums. Each term in this sum shifts the energy associated with a vertex, and as a result, the wave function coefficients on the Hartle-Hawking saddle are given by phase-weighted discrete sums over the shifted coefficients on the Einstein static universe. The latter are rational functions of the kinematic variables, and are the same as the flat spacetime coefficients known from the cosmological polytopes program, with the wave numbers replaced by the principal quantum numbers. Unlike in the Einstein static universe, the coefficients on the Hartle-Hawking saddle depend on the size of the final hypersurface $q_{1}$ through the phase $z=-\e^{2i\eta_{\rm f}}$. We identified these sums with the generalized Lauricella functions of Srivastava and Daoust at all orders. The arguments of these functions lie on the boundary of the region where they are known to converge; however, we showed that the one-site sum converges for $|\eta_{\rm f}|<\pi/2$, and that the additional total energy factor makes the two-site sum convergent. As the final hypersurface approaches the late-time conformal boundary, the one-site coefficient develops a logarithmic divergence, which is purely imaginary and does not affect the cosmological correlators.

Since shifting the external energies by two units is the same as shifting the summation variables, the Hartle-Hawking coefficients satisfy recursion relations at all orders and with any number of sites, under which all the sums collapse into rational functions of the kinematic variables. Using these relations, we derived the difference equations in the kinematic variables for the one- and two-site diagrams, which close on finite sets of two and six sums, respectively. The action of the shift operators on these sets is given by upper triangular transformation matrices, whose elements are local, rational functions of the kinematic variables, and which satisfy the consistency conditions required by the commutativity of the shifts, ensuring that the flow is independent of the path in the discrete kinematic space. This is the discrete counterpart of kinematic flow in flat FLRW spacetime. Given the initial vectors at two starting points for the one-site diagram or eight for the two-site diagram, the shift matrices determine the wave function coefficient at every point reachable by nonsingular shifts. The same strategy, where one keeps shifting the kinematic variables of the sums generated at each step until no new sums appear, can in principle be followed for diagrams with any number of sites.

Finally, we obtained the two-, three- and four-point correlation functions in terms of the wave function coefficients, where the generalized Gaunt coefficients and the angular momentum selection rules on $\mathbb{S}^3$ replace the momentum conserving Dirac delta functions of flat space. The correlators are the same on both the Hartle-Hawking saddles, and apart from the factor $1/(n_{1}n_{2}n_{3})$, the bispectrum depends on the external quantum numbers only through the total energy. At the late-time conformal boundary, all the shifted coefficients on the Einstein static universe enter the sums with the same weight, and the two-site coefficient entering the four-point function becomes real and satisfies an explicit difference equation, whose inhomogeneous terms are given by harmonic numbers of the kinematic variables.

An important direction is to formulate the generalization of these results within a consistent mathematical theory of the discrete kinematic flow. In the flat FLRW case, the twisted integrals form a basis of a finite-dimensional vector space, and the kinematic flow is described by the exterior differential calculus, with the consistency condition $\rd A+A\wedge A=0$. In the present case, since the kinematic variables are discrete, the precise language of exterior differential calculus seems to be missing. However, our results suggest that the sums play the role of the twisted integrals, the shift operators replace the exterior derivative, and the consistency condition \ref{eq:consistency} replaces $\rd A+A\wedge A=0$. Such a theory would determine, for any diagram, the number of sums that close under the shifts and the corresponding transformation matrices from a simple set of rules, without evaluating the sums explicitly. Moreover, since the wave function coefficients also depend continuously on the size of the final hypersurface through $z$, such a theory should combine the discrete shifts of the kinematic variables with the continuous dependence on $q_{1}$. Since at the late-time conformal boundary the sums are discrete counterparts of the integral representation in de Sitter spacetime with flat slicing, it would also be interesting to understand in detail how the discrete kinematic flow reduces to the continuous kinematic flow of the flat FLRW case for large principal quantum numbers, where the curvature of the spatial slices becomes negligible.

There are also several other directions in which this work can be extended. Following the strategy outlined at the end of \ref{sec:kinematic-flow}, it would be natural to derive the discrete kinematic flow explicitly for diagrams with more sites. It would also be interesting to relax the conformal coupling and to include massive fields, as well as to study loop diagrams, where the internal lines carry additional sums over the principal quantum numbers. Further, it would be interesting to understand how the correlators depend on the initial conditions for the gravitational path integral, such as the Robin conditions, the tunneling proposal, or the more recent prescriptions with regulators or a small initial boundary, and to compare our results with the scattering amplitudes in global de Sitter spacetime \cite{Taylor:2024vdc,Taylor:2025spp,Taylor:2026ase}. We leave these questions for future work.

\begin{acknowledgments}
VM thanks Paolo Creminelli, and Guilherme Pimentel for insightful discussions and helpful suggestions. The work of VM was supported by JSPS KAKENHI Grant Number 25KF0221. The work of SM was supported in part by Japan Society for the Promotion of Science Grants-in-Aid for Scientific Research No. 24K07017 and the World Premier International Research Center Initiative (WPI), MEXT, Japan. Claude Opus 5.5 (Anthropic) and ChatGPT Astra 6 (OpenAI) were used to assist with final cross-checks of calculations, preparing TikZ figures, and refining the language of the manuscript. All AI-assisted outputs were critically reviewed and independently verified by the authors, who take full responsibility for the accuracy and content of this work.
\end{acknowledgments}

\appendix
\labelformat{section}{Appendix #1}
\labelformat{subsection}{Appendix \thesection.#1}

\section{Background equations for the Hartle-Hawking saddles}
\label{app:saddles}
In this appendix, we summarize the background equations for the metric \ref{eq:metric} discussed in \ref{sec:saddles}.

The dynamics of classical gravity is governed by the Einstein-Hilbert action with cosmological constant
\begin{align}
    S = \frac{1}{2\kappa^2}\int_{\mathscr{M}} \rd^4 x\sqrt{-g} \left(R-2\Lambda\right) + S_{\partial\mathscr{M}}.
\end{align}
Here, $g_{\mu\nu}$ is the pseudo-Riemannian metric tensor with signature $(-+++)$ on the 4-manifold $\mathscr{M}$, $g$ denoting its determinant. $R$ is the Ricci scalar, $\Lambda>0$ is the cosmological constant, and we have defined $\kappa^2\equiv8\pi G$. $S_{\partial\mathscr{M}}$ is a suitable boundary term on $\partial\mathscr{M}$ required to make the variational principle consistent for given boundary conditions. For example, a variational principle consistent with Dirichlet boundary conditions requires the Gibbons-Hawking-York (GHY) boundary term \cite{York:1972sj,gibbons_action_1977,york_boundary_1986}
\begin{align}\label{eq:boundary-action}
    S_{\partial\mathscr{M}} = - \frac{1}{\kappa^2} \int_{\partial \mathscr{M}} \mathrm{d}^3y \sqrt{g^{(3)}} \, \mathscr{K},
\end{align}
where $g^{(3)}_{ij}$ is the induced metric on the boundary 3-hypersurfaces $\partial\mathscr{M}$ and $g^{(3)}$ is its determinant. $\mathscr{K}$ is the trace of the extrinsic curvature, which is defined, for example, for a constant time hypersurface, as {\sloppy $\mathscr{K}_{ij}=\frac{\sqrt{-g^{00}}}{2}\left(\partial_{t}g^{(3)}_{ij} - {}^{(3)}\nabla_{i}N_{j} - {}^{(3)}\nabla_{j}N_{i}\right)$}, here ${}^{(3)}\nabla_{i}$ is the metric compatible covariant derivative on the 3-hypersurfaces and $N^i$ is the shift vector in the standard ADM formalism. 
In the present case, for the metrics of the form \ref{eq:metric}, the shift vector vanishes.

For the metric \ref{eq:metric}, the Einstein-Hilbert action with the GHY boundary term reads as follows
\begin{equation}
S_{\text{E-H}}+S_{\partial\mathscr{M}}= 2 \pi^2 \int_0^1 \left[-\frac{3}{4N} \dot{q}^2 + N(3-\Lambda q)\right]\mathrm{d}t.
\label{bact}
\end{equation}
For an initial Neumann and final Dirichlet condition, one has to add to the above action an initial boundary term at the initial hypersurface $t=0$, that is $S_{B}=-2\pi^2\frac{3}{2N}q(0)\dot{q}(0)$ for consistency (see \cite{di2019noprescription} for an explanation). The equations of motion, with respect to $q$ and $N$ are
\begin{subequations}
\begin{align}
    \ddot{q} & = \frac{2\Lambda}{3}N^2, \label{eq:dynamical}\\
    \frac{3}{4}\dot{q}^2 & = N^2 \left(\Lambda q - 3\right). \label{eq:constraint}
\end{align}
\end{subequations}
In the above, we have also chosen the gauge fixing condition $\dot{N}=0$. Then the solution of the $q$ equation of motion \ref{eq:dynamical}, is given by
\begin{align}
    \bar{q}(t) & = \frac{\Lambda}{3} N^2 t^2 + 2 N \pi_{0} t + q_{1} - \frac{\Lambda}{3} N^2 - 2 N \pi_{0}.
\end{align}
The above solution satisfies the initial Neumann and final Dirichlet boundary conditions
\begin{align}
\bar{q}(1) = q_{1}, ~~~~~~ \frac{\dot{\bar{q}}(0)}{2N} = \pi_{0}.
\end{align}

\section{Propagator and time integrals on the complex contour}
\label{app:contour}
In this appendix, we construct the bulk-to-bulk propagator introduced in \ref{sec:propagators}, and evaluate the time integrals for the reduced wave function coefficients on the Einstein static universe discussed in \ref{sec:ESU}, along the complex contour of the conformal time.

\subsection{Bulk-to-bulk propagator}
\label{app:propagators}
Again we can write this propagator in the Fourier or Harmonic space as
\begin{align}
    G(\boldsymbol{x}_{1},\eta_{1},\boldsymbol{x}_{2},\eta_{2}) & = \sum_{n\ell m} G_{n\ell m}(\eta_{1},\eta_{2}) Q_{n\ell m}(\boldsymbol{x}_{1}) Q_{n\ell m}^*(\boldsymbol{x}_{2}) .
\end{align}
Since $Q_{n\ell m}$ are the eigenfunctions of the Laplacian on $\mathbb{S}^3$, we can define the action of the differential operator corresponding to the equation of motion of the free theory as a time dependent operator depending on the eigenvalues of the particular mode as
\begin{align}
    \mathcal{O}(\boldsymbol{x}_{1},\eta_{1})A(\eta_{1}) Q_{n\ell m}(\boldsymbol{x}_{1}) & = \mathcal{O}_{n\ell m}(\eta_{1})A(\eta_{1}) Q_{n\ell m}(\boldsymbol{x}_{1}) ,
\end{align}
Then we conclude, the differential operator evaluated at the particular eigenvalues acts on the propagator as
\begin{align}\label{eq:propagator-diff-equation}
    \mathcal{O}_{n\ell m}(\eta_{1}) G_{n\ell m}(\eta_{1},\eta_{2}) = \delta(\eta_{1}-\eta_{2}) ,
\end{align}
along with the boundary conditions
\begin{align}\label{eq:propagator-boundary}
    \lim_{\eta_{1,2}\to +i\infty+\frac{\eta_{\rm f}}{2}, \eta_{\rm f}} G_{n\ell m}(\eta_{1},\eta_{2}) = 0
\end{align}
Now we can construct this propagator using the solutions, $\varphi_{n\ell m}^{1,2}(\eta)$ to the free equation of motion as
\begin{align}
    G_{n\ell m}(\eta,\eta') = \begin{cases}
        \frac{\varphi_{n\ell m}^{1}(\eta')\varphi_{n\ell m}^{2}(\eta)}{W[\varphi_{n\ell m}^{1},\varphi_{n\ell m}^{2}]}, & \eta \succ \eta' \\
        \frac{\varphi_{n\ell m}^{1}(\eta)\varphi_{n\ell m}^{2}(\eta')}{W[\varphi_{n\ell m}^{1},\varphi_{n\ell m}^{2}]}, & \eta \prec \eta'
    \end{cases}
\end{align}
where the solutions are such that $\varphi^{1}_{n\ell m}(+i\infty+\eta_{\rm f}/2) = 0$, and $\varphi^{2}_{n\ell m}(\eta_{\rm f}) = 0$ and $W$ is the Wronskian. Moreover, the symbol $\eta_{1} \succ \eta_{2}$ denotes that $\eta_{1}$ succeeds $\eta_{2}$ on the complex conformal time contour while moving from far past ($\eta=+i\infty+\eta_{\rm f}/2$) towards the final time ($\eta_{\rm f}$), and $\eta_{1} \prec \eta_{2}$ denotes that $\eta_{1}$ precedes $\eta_{2}$. The above propagator satisfies the properties \ref{eq:propagator-diff-equation}, \ref{eq:propagator-boundary} by construction.

The particular solutions which satisfy the required boundary conditions are
\begin{subequations}
    \begin{align}
        \varphi^{1}_{n\ell m}(\eta) & = \e^{i n \eta} , \\
        \varphi^{2}_{n\ell m}(\eta) & = \e^{i n \eta} - \e^{2in\eta_{\rm f}}\e^{-i n \eta} .
    \end{align}
\end{subequations}
It is then easy to compute the Wronskian as
\begin{align}
    W[\varphi^1,\varphi^2] = \varphi^{1}_{n\ell m} ({\varphi}^{2}_{n\ell m})' - \varphi^{2}_{n\ell m} ({\varphi}^{1}_{n\ell m})' = 2in \e^{2in \eta_{\rm f}} .
\end{align}
Then the bulk-to-bulk propagator can be written as
\begin{align}
    G_{n\ell m}(\eta, \eta') = & \frac{i}{2n}\left[\e^{in(\eta'-\eta)}\Theta(\eta ; \eta') + \e^{in(\eta-\eta')}\Theta(\eta' ; \eta) \right. \nonumber\\
    & \left. -\e^{in(\eta+\eta')-2in\eta_{\rm f}} \right] .
\end{align}
Here the Heaviside theta function has the meaning
\begin{align}\label{eq:heaviside}
    \Theta(\eta_{1} ; \eta_{2}) = \begin{cases}
        1, & \eta_{1} \succ \eta_{2} \\
        0, & \eta_{1} \prec \eta_{2} \\
        \frac{1}{2}, & \eta_1 = \eta_2
    \end{cases} 
\end{align}
such that it is non-zero only when $\eta_{1}$ succeeds $\eta_{2}$ on the contour $\mathcal{C}$ running from $+i\infty+\eta_{\rm f}/2$ towards $\eta_{\rm f}$, or if the times coincide $\eta_{1}=\eta_{2}$. This definition of the Heaviside function allows us to impose correct boundary condition for the propagator. For example, say, we consider $G_{n\ell m}(+i\infty+\eta_{\rm f}/2,\eta')$ where $\eta'$ is any conformal time towards the future from the initial hypersurface. Then the first term inside the parenthesis does not contribute due to the theta function, the second term drops because it corresponds to the correct vacuum mode function, and the third term also vanishes for the same reason. On the other hand, if we consider $G_{n\ell m}(\eta_{\rm f},\eta')$, where $\eta'$ is any time in the past compared to $\eta_{\rm f}$, then the second Heaviside term drops out and the first and third terms cancel each other, satisfying the boundary condition for the propagator. Similar logic holds for the other arguments of the propagator as well.

\subsection{Time integrals in the Einstein static universe}
\label{app:ESU-integrals}

\begin{figure}[ht!]
    \centering
\begin{tikzpicture}[line width=0.8pt, >=stealth, font=\footnotesize,
    mid arrow/.style={postaction={decorate}, decoration={markings, mark=at position 0.5 with {\arrow{>}}}}]
\draw[->, lightgray] (-0.2,0) -- (3.1,0);
\draw[->, lightgray] (0,-0.2) -- (0,2.45);
\draw (2.65,2.15) -- (2.65,1.9) -- (2.95,1.9);
\node[anchor=south west, inner sep=1pt] at (2.67,1.9) {$\eta$};
\draw[gray, dashed, line width=0.6pt] (1.2,0) -- (1.2,1.6);
\node[below] at (1.2,-0.02) {$\frac{\eta_{\rm f}}{2}$};
\draw[postaction={decorate}, decoration={markings, mark=at position 0.62 with {\arrow{>}}}] (1.2,2.35) .. controls (1.2,0.75) and (1.5,0.2) .. (2.6,0);
\path (1.2,2.35) .. controls (1.2,0.75) and (1.5,0.2) .. (2.6,0) node[pos=0.3] (e2) {} node[pos=0.75] (e1) {};
\draw[fill=black] (e2.center) circle (0.045) node[right] {$\eta_{2}$};
\draw[fill=black] (e1.center) circle (0.045) node[above right, inner sep=1.5pt] {$\eta_{1}$};
\draw[fill=black] (2.6,0) circle (0.045);
\node[below] at (2.6,-0.02) {$\eta_{\rm f}$};
\node at (2.2,1.62) {$\eta_{1}\succ\eta_{2}$};
\begin{scope}[xshift=4.15cm]
\draw[->, lightgray] (-0.2,0) -- (3.55,0);
\draw[->, lightgray] (0,-0.2) -- (0,2.45);
\draw (3.1,2.15) -- (3.1,1.9) -- (3.4,1.9);
\node[anchor=south west, inner sep=1pt] at (3.12,1.9) {$\tau$};
\node[anchor=west] at (0.15,2.05) {$i\tau=\eta-\eta_{\rm f}$};
\draw[gray, dashed, line width=0.6pt] (0,1.2) -- (1.3,1.2);
\node[left] at (0,1.2) {$\frac{i\eta_{\rm f}}{2}$};
\draw[mid arrow] (3.0,1.2) -- (1.6,1.2) .. controls (0.8,1.18) and (0.3,0.75) .. (0,0);
\path (3.0,1.2) -- (1.6,1.2) node[pos=0.7] (t2) {};
\path (1.6,1.2) .. controls (0.8,1.18) and (0.3,0.75) .. (0,0) node[pos=0.45] (t1) {};
\draw[gray, dotted, line width=0.5pt] (t1.center) -- (t1.center |- 0,0);
\draw[gray, dotted, line width=0.5pt] (t2.center) -- (t2.center |- 0,0);
\draw[fill=black] (t2.center) circle (0.045) node[above, inner sep=2pt] {$\tau_{2}$};
\draw[fill=black] (t1.center) circle (0.045) node[below right, inner sep=1.5pt] {$\tau_{1}$};
\node[above] at (2.65,1.2) {$\tilde{\mathcal{C}}$};
\draw[dotted, mid arrow] (3.0,1.2) -- (3.0,0);
\node[right] at (3.0,0.6) {$\tilde{\mathcal{C}}_{1}$};
\draw[line width=1.4pt, mid arrow] (3.0,0) -- (0,0);
\node[above] at (2.6,0.02) {$\tilde{\mathcal{C}}_{2}$};
\draw[fill=black] (t1.center |- 0,0) circle (0.045);
\draw[fill=black] (t2.center |- 0,0) circle (0.045);
\node[below] at (t1.center |- 0,0) {${\rm Re}(\tau_{1})$};
\node[below] at (t2.center |- 0,0) {${\rm Re}(\tau_{2})$};
\draw[fill=black] (0,0) circle (0.045);
\end{scope}
\node at (3.9,-0.9) {$\Theta(\eta_{1};\eta_{2}) ~\longrightarrow~ \Theta\big({\rm Re}(\tau_{2})-{\rm Re}(\tau_{1})\big)$};
\end{tikzpicture}
    \caption{The conformal time $\eta$ (left) is changed to a new variable $\tau=-i(\eta-\eta_{\rm f})$ (right), which runs along a contour from $+\infty+i\eta_{\rm f}/2$ to $0$. This diagram shows the case for $N_{+}$ saddle, where $\eta_{\rm f}$ is positive. For $N_{-}$ saddle, the contour in the $\tau$-plane is reflected across the real axis. Since the integrand vanishes on the segment at infinity, for both saddles, the integral is to be performed along the real axis from $+\infty$ to $0$, and the Heaviside functions on the contour $\Theta(\eta_{1};\eta_{2})$ are replaced by $\Theta({\rm Re}(\tau_{2})-{\rm Re}(\tau_{1}))$.}
    \label{fig:conformal-time-deform}
\end{figure}
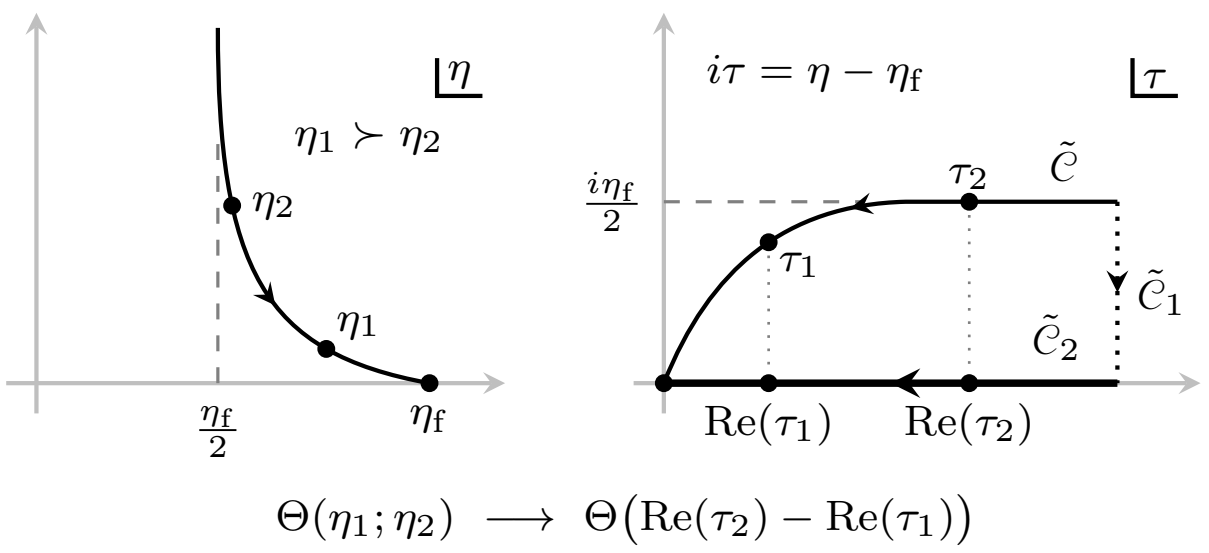

Since $\eta$ runs along the complex contour $\mathcal{C}(+i\infty+\frac{\eta_{\rm f}}{2},\eta_{\rm f})$, and the bulk-to-bulk propagator $G_{\boldsymbol{{\sf n}}}(\eta,\eta')$ contains Heaviside theta functions on the contour as defined in \ref{eq:heaviside}, to systematically perform these bulk time integrals, it is convenient to choose a new time variable $\tau=-i(\eta-\eta_{\rm f})$, such that the new contour is $\tilde{\mathcal{C}}(+\infty+\frac{i\eta_{\rm f}}{2},0)$. This new contour can be deformed into a vertical segment at infinity and a contour of integration along the real line $\tilde{\mathcal{C}}_{1}(+\infty+\frac{i\eta_{\rm f}}{2},+\infty+i0)\cup \tilde{\mathcal{C}}_{2}(+\infty,0)$. Since on the segment at infinity, the integrand is entire (assuming $\eta_{\rm f}<\pi/2$) and decays exponentially only the integration on the real line is left. Also, note that this transformation implies the Heaviside functions $\Theta(\eta_{1};\eta_{2})$ should be replaced with the conventional Heaviside function on the real line $\Theta({\rm Re}(\tau_{2})-{\rm Re}(\tau_{1}))$, see \ref{fig:conformal-time-deform}.

Then the integrals are performed as follows.
\begin{widetext}
For the one- and two- site diagrams, the integrals give:
\begin{align}
    \tilde{\psi}^{(1)}_{\rm ESU} = i\int_{\mathcal{C}(+i\infty+\frac{\eta_{\rm f}}{2},\eta_{\rm f})}\rd \eta \, \e^{i (\eta-\eta_{\rm f})\sum_{i=1}^{3} n_{i} } = -\int_{+\infty}^{0} \rd \tau_{R} \, \e^{-\tau_{R}\sum_{i=1}^{3} n_{i} } = \frac{1}{n_{1}+n_{2}+n_{3}}.
\end{align}
Here, $\tau_{R}$ is the real part of $\tau$.
\begin{align}
    \tilde{\psi}_{\rm ESU}^{(2)} & = -\int_{\mathcal{C}(+i\infty+\frac{\eta_{\rm f}}{2},\eta_{\rm f})}\rd \eta \int_{\mathcal{C}(+i\infty+\frac{\eta_{\rm f}}{2},\eta_{\rm f})}\rd \eta' K_{\boldsymbol{{\sf n}}_{1}}(\eta) K_{\boldsymbol{{\sf n}}_{2}}(\eta) (-iG_{\boldsymbol{{\sf n}}}(\eta,\eta'))K_{\boldsymbol{{\sf n}}_{3}}(\eta')K_{\boldsymbol{{\sf n}}_{4}}(\eta') \nonumber\\
    & = -\int\int_{\mathcal{C}(+i\infty+\frac{\eta_{\rm f}}{2},\eta_{\rm f})}\rd \eta \, \rd \eta' \, \e^{i(n_{1}+n_{2})(\eta-\eta_{\rm f})} \frac{1}{2n}\left[\e^{in(\eta'-\eta)}\Theta(\eta ; \eta') + \e^{in(\eta-\eta')}\Theta(\eta' ; \eta)-\e^{in(\eta+\eta')-2in\eta_{\rm f}} \right] \e^{i(n_{3}+n_{4})(\eta'-\eta_{\rm f})} \nonumber\\
    & = \int_{+\infty}^{0}\int_{+\infty}^{0}\rd \tau_{R} \, \rd \tau_{R}' \, \e^{-(n_{1}+n_{2})\tau_{R}} \frac{1}{2n}\left[\e^{-n(\tau_{R}'-\tau_{R})}\Theta(\tau_{R}'-\tau_{R}) + \e^{-n(\tau_{R}-\tau_{R}')}\Theta(\tau_{R}-\tau_{R}')-\e^{-n(\tau_{R}+\tau_{R}')} \right] \e^{-(n_{3}+n_{4})\tau_{R}'} \nonumber\\
    & = \frac{1}{(n_{1}+n_{2}+n_{3}+n_{4})(n+n_{1}+n_{2})(n+n_{3}+n_{4})} ,
\end{align}
Similarly, for the three-site diagram, one performs the following integrals
\begin{center}
    \raisebox{-26pt}{
\begin{tikzpicture}[line width=1.0pt, scale=2]
\draw (0,0) -- (1,0);
\draw (1,0) -- (2,0);
\draw[lightgray,line width=1.0pt] (0,0) -- (-0.25,0.55);
\draw[lightgray,line width=1.0pt] (0,0) -- (0.25,0.55);
\draw[lightgray,line width=1.0pt] (1,0) -- (1.0,0.55);
\draw[lightgray,line width=1.0pt] (2,0) -- (1.75,0.55);
\draw[lightgray,line width=1.0pt] (2,0) -- (2.25,0.55);
\node[scale=1,above] at (-0.25,0.55) {$n_{1}$};
\node[scale=1,above] at (0.25,0.55) {$n_{2}$};
\node[scale=1,above] at (1.0,0.55) {$n_{3}$};
\node[scale=1,above] at (1.75,0.55) {$n_{4}$};
\node[scale=1,above] at (2.25,0.55) {$n_{5}$};
\draw[lightgray, line width=2.0pt] (-0.6,0.55) -- (2.6,0.55);
\draw[fill=black] (0,0) circle (.03cm);
\draw[fill=black] (1,0) circle (.03cm);
\draw[fill=black] (2,0) circle (.03cm);
\node[scale=1] at (0.5,-.15) {$n_a$};
\node[scale=1] at (1.5,-.15) {$n_b$};
\end{tikzpicture}
}
\end{center}
\begin{align}
    \tilde{\psi}_{\rm ESU}^{(3)} & = -i\int\int\int_{\mathcal{C}(+i\infty+\frac{\eta_{\rm f}}{2},\eta_{\rm f})}\rd \eta \rd \eta' \rd \eta'' K_{\boldsymbol{{\sf n}}_{1}}(\eta) K_{\boldsymbol{{\sf n}}_{2}}(\eta) (-iG_{\boldsymbol{{\sf n}}_a}(\eta,\eta'))K_{\boldsymbol{{\sf n}}_{3}}(\eta')(-iG_{\boldsymbol{{\sf n}}_b}(\eta',\eta''))K_{\boldsymbol{{\sf n}}_{4}}(\eta'')K_{\boldsymbol{{\sf n}}_{5}}(\eta'') \nonumber\\
    & = -\int_{+\infty}^{0}\int_{+\infty}^{0}\int_{+\infty}^{0}\rd \tau_{R} \, \rd \tau_{R}' \, \rd \tau_{R}'' \, \e^{-(n_{1}+n_{2})\tau_{R}} \frac{1}{2n_a}\left[\e^{-n_a(\tau_{R}'-\tau_{R})}\Theta(\tau_{R}'-\tau_{R}) + \e^{-n_a(\tau_{R}-\tau_{R}')}\Theta(\tau_{R}-\tau_{R}')-\e^{-n_a(\tau_{R}+\tau_{R}')} \right] \nonumber\\
    &~~  \times \e^{-n_{3}\tau_{R}'}\frac{1}{2n_b}\left[\e^{-n_b(\tau_{R}''-\tau_{R}')}\Theta(\tau_{R}''-\tau_{R}') + \e^{-n_b(\tau_{R}'-\tau_{R}'')}\Theta(\tau_{R}'-\tau_{R}'')-\e^{-n_b(\tau_{R}'+\tau_{R}'')} \right] \e^{-(n_{4}+n_{5})\tau_{R}''} \nonumber\\
    & = \frac{n_a+n_b+n_1+n_2+2n_3+n_4+n_5}{\left(n_1+n_2+n_3+n_4+n_5\right) \left(n_a+n_1+n_2\right) \left(n_a+n_3+n_4+n_5\right) \left(n_b+n_1+n_2+n_3\right) \left(n_b+n_4+n_5\right) \left(n_a+n_b+n_3\right)} .
\end{align}
\end{widetext}

\section{Shift of $\tilde{\psi}^{(2)}_{\rm HH}$ with respect to external and internal energies}
\label{sec:kin-flow-two-site-derivation}
At $V=2$, the wave function coefficient is given by the following double sum
\begin{align}
     & \tilde{\psi}_{\rm HH}^{(2)} =\sum_{k_{1}=0}^{\infty} \sum_{k_{2}=0}^{\infty}\frac{2^2 (-1)^{k_{1}+k_{2}}\e^{i(2k_{1}+2k_{2}+2)\eta_{\rm f}}}{(X_{1}+X_{2}+2(k_{1}+k_{2})+2)} \nonumber\\
    &~~~~ \times \frac{1}{(X_{1}+2k_{1}+1+Y_{12})(X_{2} + 2 k_{2} + 1 +Y_{12})}.
\end{align}
Acting once with the sum collapsing shift operators with respect to the external energies, we get,
\begin{align}
     & (\e^{2i\eta_{\rm f}}S_{X_{1}}^{(2)}+1)\tilde{\psi}_{\rm HH}^{(2)} =\frac{2^2\e^{2i\eta_{\rm f}}}{(X_{1}+1+Y_{12})} \nonumber\\
    & \times\sum_{k_{2}=0}^{\infty}\frac{ (-1)^{k_{2}}\e^{2ik_{2}\eta_{\rm f}}}{(X_{1}+X_{2}+2k_{2}+2)(X_{2} + 2 k_{2}+1+Y_{12})} .
\end{align}
The remaining summand can be rewritten using the partial fraction relation as
\begin{align}
     & (\e^{2i\eta_{\rm f}}S_{X_{1}}^{(2)}+1)\tilde{\psi}_{\rm HH}^{(2)} =\frac{2^2\e^{2i\eta_{\rm f}}}{(X_{1}+1+Y_{12})}\sum_{k_{2}=0}^{\infty}\frac{(-1)^{k_{2}}\e^{2ik_{2}\eta_{\rm f}}}{X_{1}+1-Y_{12}} \nonumber\\
    & \times\left[\frac{1}{X_{2} + 2 k_{2}+1+Y_{12}}-\frac{1}{X_{1}+X_{2}+2k_{2}+2}\right] .
\end{align}
Due to the partial fraction, the one-dimensional sum over the two terms can now be expressed in terms of $\tilde{\psi}^{(1)}_{\rm HH}$, giving
\begin{align}
     S_{X_{1}}^{(2)}\tilde{\psi}_{\rm HH}^{(2)} = & -\e^{-2i\eta_{\rm f}}\tilde{\psi}_{\rm HH}^{(2)} + \frac{2\e^{-i\eta_{\rm f}}}{(X_{1}+1)^2-Y_{12}^2} F_{2}^{-} ,
\end{align}
where the object $F^{\pm}_{2}$ corresponds to lower order sums, and is given by
\begin{align}\label{eq:F-def2}
    F_{2}^{\pm} = \tilde{\psi}^{(1)}_{\rm HH}(X_{2} + Y_{12}) \pm \tilde{\psi}^{(1)}_{\rm HH}(X_{1}+X_{2}+1) .
\end{align}
Since $\tilde{\psi}^{(2)}_{\rm HH}$ is symmetric under the exchange of $X_{1}\leftrightarrow X_{2}$, we can also write
\begin{align}
     S_{X_{2}}^{(2)}\tilde{\psi}_{\rm HH}^{(2)} = & -\e^{-2i\eta_{\rm f}}\tilde{\psi}_{\rm HH}^{(2)} + \frac{2\e^{-i\eta_{\rm f}}}{(X_{2}+1)^2-Y_{12}^2} F^{-}_{1} ,
\end{align}
where the object $F^{\pm}_{1}$ is given by
\begin{align}\label{eq:F-def1}
    F_{1}^{\pm} = \tilde{\psi}^{(1)}_{\rm HH}(X_{1} + Y_{12}) \pm \tilde{\psi}^{(1)}_{\rm HH}(X_{1}+X_{2}+1) .
\end{align}
Now, for convenience, let us define
\begin{subequations}
\begin{align}
    B_{1} & = X_{1} + Y_{12} + 2 k_{1} + 1 , \\
    B_{2} & = X_{2} + Y_{12} + 2 k_{2} + 1 , \\
    B_{3} & = X_{1} + X_{2} + 2 (k_{1}+k_{2}) + 2 .
\end{align}
\end{subequations}
Then the 2-site wave function coefficient on Hartle-Hawking saddle is given by the sum of the form
\begin{align}
    \tilde{\psi}^{(2)}_{\rm HH}(X_{1},X_{2},Y_{12})=\sum_{k_{1},k_{2}=0}^{\infty}\frac{2^2(-1)^{k_{1}+k_{2}}\e^{i(2k_{1}+2k_{2}+2)\eta_{\rm f}}}{B_{1}(k_{1})B_{2}(k_{2})B_{3}(k_{1},k_{2})} .
\end{align}
From the definitions of $B_{1,2,3}$, we can write the following
\begin{align}
    B_{1} + B_{2} - B_{3} = 2Y_{12} .
\end{align}
This allows us to write the summand in the following partial fraction form
\begin{align}
    \frac{1}{B_{1}B_{2}B_{3}}=\frac{1}{2Y_{12}} \left[\frac{1}{B_{1}B_{3}}+\frac{1}{B_{2}B_{3}}-\frac{1}{B_{1}B_{2}}\right] ,
\end{align}
Now let us focus on sum corresponding to the first term inside the parenthesis
\begin{align}
    G_{1} & \equiv\sum_{k_{1},k_{2}=0}^{\infty} \frac{(-1)^{k_{1}+k_{2}}\e^{2i(k_{1}+k_{2})\eta_{\rm f}}}{B_{1}B_{3}} \nonumber\\
    & = \frac{\e^{-i\eta_{\rm f}}}{2}\sum_{k_{1}=0}^{\infty}\frac{(-1)^{k_{1}}\e^{2ik_{1}\eta_{\rm f}}\,\tilde{\psi}^{(1)}_{\rm HH}(X_{1}+X_{2}+2k_{1}+1)}{X_{1}+Y_{12}+2k_{1}+1} .
\end{align}
Similarly for the second term in the parenthesis we get
\begin{align}
    G_{2} & \equiv  \sum_{k_{1},k_{2}=0}^{\infty} \frac{(-1)^{k_{1}+k_{2}}\e^{2i(k_{1}+k_{2})\eta_{\rm f}}}{B_{2}B_{3}} \nonumber\\
    & = \frac{\e^{-i\eta_{\rm f}}}{2}\sum_{k_{2}=0}^{\infty}\frac{(-1)^{k_{2}}\e^{2ik_{2}\eta_{\rm f}}\,\tilde{\psi}^{(1)}_{\rm HH}(X_{1}+X_{2}+2k_{2}+1)}{X_{2}+Y_{12}+2k_{2}+1} .
\end{align}
While for the third term inside the parenthesis, the $k_{1,2}$ sums decouple and we get the product of two one-site coefficients $G$ defined in \ref{eq:G-def},
\begin{align}
    \sum_{k_{1},k_{2}=0}^{\infty} \frac{(-1)^{k_{1}+k_{2}}\e^{2i(k_{1}+k_{2})\eta_{\rm f}}}{B_{1}B_{2}} = \frac{\e^{-2i\eta_{\rm f}}}{4}\psi_{1Y}\psi_{2Y} = G .
\end{align}
Thus the 2-site wave function coefficient can be written as
\begin{align}\label{eq:psi2-partial-fraction}
   \tilde{\psi}^{(2)}_{\rm HH} =\frac{2^2\e^{2i\eta_{\rm f}}}{2Y_{12}} \left[ G_{1} + G_{2} - G \right] .
\end{align}
Since $B_{3}$ is independent of $Y_{12}$, shifting the object $G_{1}$ with respect to $Y_{12}\to Y_{12}+2$ is equivalent to shifting $k_{1}\to k_{1}+1$ in $B_{1}$ at fixed $k_{1}+k_{2}$, then using \ref{eq:sum-collapse} for the sum over $k_{1}$, partial fraction and the one-site recursion relation \ref{eq:one-site-recursion}, we get
\begin{align}
    & S_{Y_{12}}^{(2)}G_{1} = G_{1} + \frac{\upsilon}{2(X_{2}-Y_{12}-1)}\left[\frac{2\upsilon}{X_{1}+Y_{12}+1}\right. \nonumber\\
    &~~ \left. - \upsilon^{2}\psi_{1Y} - \psi_{12}\right] - \frac{\upsilon\,\psi_{12}}{2(X_{1}+Y_{12}+1)} ,
\end{align}
where, as above, $\upsilon\equiv\e^{-i\eta_{\rm f}}$. Because the wave function is symmetric under the exchange of the vertex variables, the shift of $G_2$ can be obtained directly from that of $G_1$ by exchanging $X_1 \leftrightarrow X_2$ and $\psi_{1Y} \leftrightarrow \psi_{2Y}$. Further, the one-site recursion relation \ref{eq:one-site-recursion} gives
\begin{align}
    S_{X_{1}}^{(2)}\psi_{1Y} & = S_{Y_{12}}^{(2)}\psi_{1Y} = -\upsilon^{2}\psi_{1Y} + \frac{2\upsilon}{X_{1}+Y_{12}+1} , \nonumber\\
    S_{X_{1}}^{(2)}\psi_{12} & = S_{X_{2}}^{(2)}\psi_{12} = -\upsilon^{2}\psi_{12} + \frac{2\upsilon}{X_{1}+X_{2}+2} ,
\end{align}
while $S_{X_{2}}^{(2)}\psi_{1Y}=\psi_{1Y}$ and $S_{Y_{12}}^{(2)}\psi_{12}=\psi_{12}$, and the shifts of $\psi_{2Y}$ are obtained by exchanging $X_1 \leftrightarrow X_2$. Since $G$ is a product of $\psi_{1Y}$ and $\psi_{2Y}$, its shifts follow from the above,
\begin{align}
    S_{X_{1}}^{(2)} G & = -\upsilon^{2} G + \frac{\upsilon^{3}\psi_{2Y}}{2(X_{1}+Y_{12}+1)} , \nonumber\\
    S_{Y_{12}}^{(2)} G & = \upsilon^{4} G - \frac{\upsilon^{5}\psi_{1Y}}{2(X_{2}+Y_{12}+1)} - \frac{\upsilon^{5}\psi_{2Y}}{2(X_{1}+Y_{12}+1)} \nonumber\\
    & ~~~~ + \frac{\upsilon^{4}}{(X_{1}+Y_{12}+1)(X_{2}+Y_{12}+1)} ,
\end{align}
and $S_{X_{2}}^{(2)}G$ is obtained by exchanging $X_1 \leftrightarrow X_2$ and $\psi_{1Y} \leftrightarrow \psi_{2Y}$. Thus, shifting \ref{eq:psi2-partial-fraction} with respect to $Y_{12}$ and using it again to eliminate $G_{1}+G_{2}$, the shift of $\tilde{\psi}^{(2)}_{\rm HH}$ with respect to $Y_{12}$ can be expressed as
\begin{align}
   & S_{Y_{12}}^{(2)}\tilde{\psi}^{(2)}_{\rm HH} = \frac{Y_{12}}{Y_{12}+2}\tilde{\psi}^{(2)}_{\rm HH} + \frac{2^2\e^{2i\eta_{\rm f}}}{2Y_{12}+4} \Big[ G - S_{Y_{12}}^{(2)}G \nonumber\\
   & ~~ + S_{Y_{12}}^{(2)}G_{1} - G_{1} + S_{Y_{12}}^{(2)}G_{2} - G_{2} \Big] .
\end{align}
Therefore, we have shown that the shifts $S^{(2)}_{I}$ of set of sums $(\tilde{\psi}^{(2)}_{\rm HH},\psi_{1Y},\psi_{2Y},\psi_{12},G,1)$ with respect to the internal and external energies closes.
 
\subsection{Transformation matrices}
\label{app:two-site-matrices}
Collecting the above shifts, the sub-blocks of the transformation matrix $\mathscr{A}^{(2)}_{K}$ in \ref{eq:block-matrix} can be read off as follows. For convenience let us write $\upsilon \equiv \e^{-i\eta_{\rm f}}$ ($z=-1/\upsilon^2$), then for the shift with respect to the external energy $X_1$, the non-zero sub-blocks evaluate to:
\begin{align}
    M_{\Psi\Psi}^{X_1} &= -\upsilon^{2}, \quad M_{\Psi F}^{X_1} = \begin{pmatrix} 0 & \upsilon c_{X_1} & -\upsilon c_{X_1} \end{pmatrix} , \nonumber\\
    M_{GG}^{X_1} &= -\upsilon^{2}, \quad M_{GF}^{X_1} = \frac{\upsilon^{3}}{4}\begin{pmatrix} 0 & s_1 & 0 \end{pmatrix} , \nonumber\\
    M_{FF}^{X_1} &= \mathrm{diag}(-\upsilon^{2}, 1, -\upsilon^{2}), \quad
    M_{FR}^{X_1} = \begin{pmatrix} \upsilon s_1 \\ 0 \\ \upsilon t_{12} \end{pmatrix} ,
\end{align}
where we have defined
\begin{align*}
    c_{X_i} &= \frac{2}{(X_i+1)^2-Y_{12}^2} , \quad
    s_i = \frac{2}{X_i+Y_{12}+1} , \\
    t_{12} &= \frac{2}{X_1+X_2+2} .
\end{align*}
Similarly, for the shifts with respect to $X_{2}$ the matrix elements take the form
\begin{align}
    M_{\Psi\Psi}^{X_2} &= -\upsilon^{2}, \quad M_{\Psi F}^{X_2} = \begin{pmatrix} \upsilon c_{X_2} & 0 & -\upsilon c_{X_2} \end{pmatrix} , \nonumber\\
    M_{GG}^{X_2} &= -\upsilon^{2}, \quad M_{GF}^{X_2} = \frac{\upsilon^{3}}{4}\begin{pmatrix} s_2 & 0 & 0 \end{pmatrix} , \nonumber\\
    M_{FF}^{X_2} &= \mathrm{diag}(1, -\upsilon^{2}, -\upsilon^{2}), \quad
    M_{FR}^{X_2} = \begin{pmatrix} 0 \\ \upsilon s_2 \\ \upsilon t_{12} \end{pmatrix} .
\end{align}
And finally, for the shifts with respect to internal energies, the matrix elements take the following form
\begin{align}
    M_{\Psi\Psi}^{Y_{12}} &= 1-c_{Y} , \quad M_{\Psi G}^{Y_{12}} = c_{Y}\left(\upsilon^{-2}-\upsilon^{2}\right) , \nonumber\\
    M_{\Psi F}^{Y_{12}} &= c_{Y}\begin{pmatrix} \upsilon^{3}s_{2}/4-\upsilon b_{2} \\ \upsilon^{3}s_{1}/4-\upsilon b_{1} \\ -\upsilon^{-1}\left(b_{1}+b_{2}+(s_{1}+s_{2})/4\right) \end{pmatrix}^{T} , \nonumber\\
    M_{\Psi R}^{Y_{12}} &= c_{Y}\left(b_{2}s_{1}+b_{1}s_{2}-\frac{\upsilon^{2}}{4}s_{1}s_{2}\right) , \nonumber\\
    M_{GG}^{Y_{12}} &= \upsilon^{4} , \quad M_{GF}^{Y_{12}} = -\frac{\upsilon^{5}}{4}\begin{pmatrix} s_{2} & s_{1} & 0 \end{pmatrix} , \nonumber\\
    M_{GR}^{Y_{12}} &= \frac{\upsilon^{4}}{4}s_{1}s_{2} , \quad
    M_{FF}^{Y_{12}} = \mathrm{diag}(-\upsilon^{2}, -\upsilon^{2}, 1) , \nonumber\\
    M_{FR}^{Y_{12}} &= \begin{pmatrix} \upsilon s_1 \\ \upsilon s_2 \\ 0 \end{pmatrix} ,
\end{align}
where we have used the following definitions
\begin{align*}
    c_Y & = \frac{2}{Y_{12}+2} , \qquad b_{i} = \frac{1}{2(X_i-Y_{12}-1)} .
\end{align*}

\subsection{Late-time difference equation}
\label{app:late-time}
In this appendix, we prove the late-time difference equation \ref{eq:late-time-psi2} and the single-sum representation \ref{eq:late-time-single-sum} of the two-site coefficient.

For real $\eta_{\rm f}$, the absolute value of the summand of the double sum for $\tilde{\psi}^{(2)}_{\rm HH}$, given by \ref{eq:HH-sum-general} with $V=2$, is $4\tilde{\psi}^{(2)}_{\rm ESU}(X_{1}+2k_{1}+1,X_{2}+2k_{2}+1,Y_{12})$, which does not depend on $\eta_{\rm f}$. Since all the kinematic variables are positive, and $k_{1}+k_{2}+2\geq2\sqrt{(k_{1}+1)(k_{2}+1)}$, this is bounded as
\begin{align}
    & \tilde{\psi}^{(2)}_{\rm ESU}(X_{1}+2k_{1}+1,X_{2}+2k_{2}+1,Y_{12}) \nonumber\\
    & ~~~~ \leq \frac{1}{(k_{1}+k_{2}+2)(k_{1}+1)(k_{2}+1)} \nonumber\\
    & ~~~~ \leq \frac{1}{2(k_{1}+1)^{3/2}(k_{2}+1)^{3/2}} ,
\end{align}
whose double sum converges. Therefore, the double sum converges absolutely and uniformly in $\eta_{\rm f}$, and $\tilde{\psi}^{(2)}_{\rm HH}$ is a continuous function of $\eta_{\rm f}$ up to the late-time conformal boundary. As a result, its late-time limit is obtained by setting $\eta_{\rm f}=\pm\pi/2$ in the summand, where $2(-1)^{k}\e^{i(2k+1)\eta_{\rm f}}=\pm2i$ for every $k$, which gives \ref{eq:psi2-late}. In particular, $\tilde{\psi}^{(2)}_{\rm late}$ is real, and the same on both the saddles.

Shifting $X_{1}\to X_{1}+2$ in \ref{eq:psi2-late} is the same as shifting $k_{1}\to k_{1}+1$, so that, owing to the absolute convergence, all the terms cancel except the ones with $k_{1}=0$,
\begin{align}
    & \tilde{\psi}^{(2)}_{\rm late}(X_{1}+2,X_{2},Y_{12}) - \tilde{\psi}^{(2)}_{\rm late}(X_{1},X_{2},Y_{12}) \nonumber\\
    & ~~~~ = 4\sum_{k_{2}=0}^{\infty}\tilde{\psi}^{(2)}_{\rm ESU}(X_{1}+1,X_{2}+2k_{2}+1,Y_{12}) \nonumber\\
    & ~~~~ = \frac{4}{X_{1}+Y_{12}+1}\sum_{k_{2}=0}^{\infty}\frac{1}{B_{2}B_{3}}\bigg|_{k_{1}=0} ,
\end{align}
with $B_{2}$ and $B_{3}$ defined above. Since $B_{3}-B_{2}=X_{1}-Y_{12}+1$ at $k_{1}=0$, the remaining summand can be split using the partial fraction
\begin{align}
    \frac{1}{B_{2}B_{3}}\bigg|_{k_{1}=0} = \frac{1}{X_{1}-Y_{12}+1}\left[\frac{1}{B_{2}}-\frac{1}{B_{3}}\right]_{k_{1}=0} ,
\end{align}
where $X_{1}-Y_{12}+1>0$ due to the angular momentum selection rules \ref{eq:angular-momentum-selection}. Using the identity
\begin{align}
    \sum_{k=0}^{\infty}\left[\frac{1}{k+a}-\frac{1}{k+b}\right] = H_{b-1}-H_{a-1} ,
\end{align}
which follows from the definition of the harmonic numbers below \ref{eq:one-site-late-asym}, with $B_{2}=2(k_{2}+a)$ and $B_{3}|_{k_{1}=0}=2(k_{2}+b)$, that is, $a=(X_{2}+Y_{12}+1)/2$ and $b=(X_{1}+X_{2}+2)/2$, we obtain
\begin{align}
    & \tilde{\psi}^{(2)}_{\rm late}(X_{1}+2,X_{2},Y_{12}) - \tilde{\psi}^{(2)}_{\rm late}(X_{1},X_{2},Y_{12}) \nonumber\\
    & ~~~~ = \frac{2}{(X_{1}+1)^{2}-Y_{12}^{2}}\left[H_{\frac{X_{1}+X_{2}}{2}} - H_{\frac{X_{2}+Y_{12}-1}{2}}\right] ,
\end{align}
which is \ref{eq:late-time-psi2}. The shift with respect to $X_{2}$ follows from the symmetry under the exchange of the labels $1\leftrightarrow2$.

The same result follows from the first row of $\mathscr{A}^{(2)}_{X_{1}}$ in \ref{eq:two-site-late}, $S^{(2)}_{X_{1}}\tilde{\psi}^{(2)}_{\rm HH}\to\tilde{\psi}^{(2)}_{\rm HH}-ic_{X_{1}}(\psi_{2Y}-\psi_{12})$, since \ref{eq:one-site-late-asym} gives $\psi_{2Y}-\psi_{12}\to-i\left[H_{\frac{X_{2}+Y_{12}-1}{2}}-H_{\frac{X_{1}+X_{2}}{2}}\right]$ at the late-time conformal boundary of the $N_{+}$ saddle, and similarly for the $N_{-}$ saddle.

Finally, each term of the double sum in \ref{eq:psi2-late} decreases monotonically to zero as $X_{1}$ increases, and is bounded by the above $X_{1}$-independent bound, so that $\tilde{\psi}^{(2)}_{\rm late}\to0$ for $X_{1}\to\infty$. Then, adding \ref{eq:late-time-psi2} for $X_{1},X_{1}+2,\dots,X_{1}+2(m-1)$, we find
\begin{align}
    & \tilde{\psi}^{(2)}_{\rm late}(X_{1}+2m,X_{2},Y_{12}) - \tilde{\psi}^{(2)}_{\rm late}(X_{1},X_{2},Y_{12}) \nonumber\\
    & ~~~~ = \sum_{j=0}^{m-1}c_{X_{1}+2j}\left[H_{\frac{X_{1}+X_{2}}{2}+j} - H_{\frac{X_{2}+Y_{12}-1}{2}}\right] ,
\end{align}
and taking $m\to\infty$ gives \ref{eq:late-time-single-sum}.

\bibliographystyle{apsrev4-2}
\bibliography{reference}
\end{document}